\documentstyle[eqsecnum,aps,floats,epsf]{revtex}
\newcommand{\cd}{\cdot}

\newcommand{\al}{\alpha}

\def \g {\gamma}
\def \a {\alpha}

\newcommand{\de}{\delta}
\newcommand{\De}{\Delta}
\newcommand{\ep}{\epsilon}
\newcommand{\ga}{\gamma}

\newcommand{\La}{\Lambda}
\newcommand{\la}{\lambda}
\newcommand{\Om}{\Omega}
\newcommand{\om}{\omega}
\newcommand{\si}{\sigma}
\newcommand{\Si}{\Sigma}

\newcommand{\ra}{\rightarrow}
\newcommand{\DD}{\mbox{$\cal D$}}
\newcommand{\GG}{\mbox{$\cal G$}}
\newcommand{\SS}{\mbox{$\cal S$}}

\newcommand{\be}{\begin{equation}}
\newcommand{\ee}{\end{equation}}

\newcommand{\lsim}{\stackrel{<}{\sim}}
\newcommand{\bea}{\begin{eqnarray}}
\newcommand{\eea}{\end{eqnarray}}
\newcommand{\bean}{\begin{eqnarray*}}
\newcommand{\eean}{\end{eqnarray*}}
\newcommand{\dd}{\partial}

\newcommand{\bk}{{\bf k}}

\def\spose#1{\hbox to 0pt{#1\hss}} 
\def\ltapprox{\mathrel{\spose{\lower 3pt\hbox{$\mathchar"218$}} 
 \raise 2.0pt\hbox{$\mathchar"13C$}}} 
\def\gtapprox{\mathrel{\spose{\lower 3pt\hbox{$\mathchar"218$}} 
 \raise 2.0pt\hbox{$\mathchar"13E$}}} 
\def\inapprox{\mathrel{\spose{\lower 3pt\hbox{$\mathchar"218$}} 
 \raise 2.0pt\hbox{$\mathchar"232$}}}

\begin{document}
\draft
\preprint{\ 
\begin{tabular}{rr}
& 
\end{tabular}
} 
\title{ Particle Creation in Pre-Big-Bang Cosmology:\\ theory and
observational consequences}
\author{ R. Durrer$^a$, K.E. Kunze$^{a,b}$ and 
        M. Sakellariadou$^{c}$ }
\address{ 
\smallskip\smallskip
$^a$D\'epartement de Physique Th\'eorique, Universit\'e de 
Gen\`eve,\\
24 quai Ernest Ansermet, CH-1211 Gen\`eve 4, Switzerland.\\
\smallskip\smallskip
$^b$ Fakult\"at f\"ur Physik, Universit\"at Freiburg, \\ 
 Hermann Herder Strasse 3,
	D-79104 Freiburg, Germany.\\
${}^c$Department of Astrophysics, Astronomy, and Mechanics,\\
       Faculty of Physics, University of Athens,\\
       Panepistimiopolis, GR-15784 Zografos, Hellas
\\~}
\date{\today}

\maketitle

\begin{abstract}

We present some phenomenological aspects of the pre-big-bang
cosmological model inspired by the duality properties of string
theory. In particular, assuming the spatial sections of the homogeneous
background geometry to be isotropic, we discuss the quantum production
of perturbations of the background fields (gravitons, dilatons, moduli
fields), as well as the production of particles which do not
contribute to the background, which we call ``seeds''. As such we
consider the cases of electromagnetic and axionic seeds.  We also
discuss their possible observational consequences, for example,
we study whether they can provide the origin of primordial galactic
magnetic fields, and whether they can generate the initial
fluctuations leading to the formation of large-scale structure and the
measured cosmic microwave background anisotropies. We finally analyze
axion and photon production in four dimensional anisotropic
pre-big-bang cosmological models.

\end{abstract}

\newpage

\section{Introduction}

The pre-big-bang scenario (PBB) is a cosmological
model inspired by the duality properties of string
theory~\cite{pbb1,pbb2,lwc}. It differs
from conventional cosmological models based on general relativity by
the presence of the dilaton field.  In the PBB model, the initial
state of the universe is the string perturbative vacuum, instead of a
hot and dense state as predicted by the standard cosmological
model. In the string (S) frame -- in which weakly coupled strings move
along geodesic surfaces~\cite{sf} -- the low-energy string effective
action in $D$ dimensions, neglecting both finite-size effects and
 loop corrections, can be written as
\bea
\label{action}
S&=&-{1\over 16 \pi G_{\rm D}}\int d^D x ~~\sqrt{|g_{\rm D}|}
~~e^{-\phi} \nonumber \\ &\times & \left[R_{\rm D}+
\partial_\mu\phi\partial^\mu\phi - {1\over 12}H_{\mu\nu\alpha}
H^{\mu\nu\alpha}+ V(\phi)\right ]~+~S_{\rm M}~, 
\eea 
where $ G_{\rm D}$ is Newton's constant in $D$ dimensions and
$R_{\rm D}$ is the Ricci scalar of the $D$-dimensional space-time. In
Eq.~(\ref{action}), $H_{\mu\nu\alpha}$ denotes the field strength of
the two-index antisymmetric tensor $B_{\mu\nu}$ (i.e., $H=dB$), $\phi$
is the dilaton field and $V(\phi)$ is the dilaton potential.  During
the pre-big-bang era when the effective coupling $e^{\phi/2}$ is
small, it is usually assumed that the dilaton potential ($V\sim 
\exp[-\exp(-\phi)]$) can be neglected. On the other hand, during the
post-big-bang era the dilaton potential vanishes, leading to a
constant massive dilaton sitting at the minimum.  The
action $S_{\rm M}$ describes ``bulk'' string matter satisfying the
classical string equations of motion in a given background.

The low-energy string effective action, Eq.~(\ref{action}), leads to a
set of cosmological equations for a homogeneous and spatially flat
background which obey the ``scale-factor duality''~\cite{pbb1}: if the
background fields are only time-dependent, and if $\{a_i(t),\phi(t);
i=1,...,D-1\}$ denote the scale factors and dilaton of a given
homogeneous exact solution with vanishing antisymmetric tensor, then
the system $\{\tilde{a}_i(t),\tilde{\phi}(t); i=1,...,D-1\}$, obtained
through the transformation in $(D-1)$ spatial dimensions:
\be
a_i\rightarrow \tilde{a}_i=a_i^{-1}~,~\phi\rightarrow
\tilde{\phi}=\phi-2\sum_i\ln a_i~, 
\ee 
is a new exact solution of the cosmological equations~\cite{pbb1,dt}.
In the presence of matter, which can be approximated as a perfect
fluid, a scale factor duality remains a symmetry of the solutions if
we add the transformation law $p/\rho \rightarrow
-p/\rho$~\cite{pbb1}.

The solution of the string cosmology equations with the perturbative
vacuum as initial condition, describes a phase of growing curvature
and growing dilaton. It is called the ``dilaton phase''. Once the
curvature $H^2$ (where $H=\dot a/a$; a dot stands for differentiation
with respect to cosmic time $t$) reaches the string scale
$\lambda_{\rm s}^{-1}$, the background enters the ``string phase''
during which  higher derivative terms in the $\alpha'$ expansion
become important.  During this phase, the curvature scale is supposed
to remain constant while the string coupling increases.  At the
beginning of the string phase, the string coupling can be arbitrarily
small; this is one of the parameters of the model.  At the end of the
string phase, the dilaton approaches the strong coupling regime (the
coupling constant is $\sim 1$), leading to the transition to the
radiation-dominated era characterized by  decelerated evolution and 
frozen dilaton field. The duration of the string phase is the second
parameter of this cosmological model inspired by string theory.  A
complete description of the high curvature strong coupling regime
and its transition to a radiation-dominated post-big-bang universe
with frozen dilaton is still lacking. Some ``toy models'' and
approximations have however been studied~\cite{vb,cc}.

By combining the duality transformation with time-reversal symmetry
($t\rightarrow -t$), another symmetry of the model, we can always
associate to any given decelerated expanding post-big-bang solution
with decreasing curvature:
\be 
\dot a > 0~~,~~\ddot a < 0~~, ~~\dot H < 0~, 
\ee 
an accelerated expanding pre-big-bang solution with growing curvature:
\be 
\dot a > 0~~, ~~\ddot a > 0~~, ~~\dot H > 0~.  
\ee
Transforming this accelerating solution into the Einstein (E) frame --- in
which the dilaton coupling is absorbed into the metric by a conformal
transformation, i.e., the graviton and dilaton kinetic terms are
diagonalized --- through the conformal $(D-1)$-dimensional
rescaling~\cite{pbb2}:
\be 
a\rightarrow a ~e^{-\phi/(D-2)} ~~~,~~~
dt\rightarrow dt ~e^{-\phi/(D-2)}~, 
\ee 
the pre-big-bang solution transforms into a Friedmann universe with
accelerated contraction and growing curvature
\be 
\dot a < 0~~, ~~\ddot a < 0~~, ~~\dot H < 0~.
\ee 
Let us consider as an example~\cite{pbb1}, a spatially flat
isotropic background in $D=4$ space-time dimensions with vanishing
antisymmetric tensor and dilaton potential and with non-zero bulk
string matter. The scale factor duality maps the standard
radiation-dominated solution with frozen dilaton
\be
a\sim t^{1/2}~~,~~\phi={\rm const.}~~,~~\rho=3 p~~,~~t>0 
\ee 
into the pre-big-bang solution 
\be 
a\sim (-t)^{-1/2}~,~\phi=-3\ln(-t)~,~\rho=-3 p~~,~~t<0
\ee
which describes an accelerated expansion of the pole- or
super-inflation type~\cite{pole} with growing curvature and
gravitational coupling.  

The phenomenological drawbacks (i.e.,
horizon, flatness problems) of the standard cosmological model can be
equally well solved, either by a sufficiently long era of
super-inflation in the string frame, or an era of accelerated
contraction in the Einstein frame.  Thus, the choice of frame is just
a matter of convenience, and the physical description of the PBB
scenario is equivalent in the S and E frames~\cite{pbb2}. In this
paper we usually work in the string frame, otherwise we specify it.

The usual relations with the adiabatic decrease of temperature $T$:
\be
\rho \sim 1/a^{\rm D} \sim T^{\rm D} ~~~,~~~ T\sim 1/a
\ee
in the radiation-dominated background, are mapped~\cite{pbb2} to:
\be
\rho \sim 1/a^{{\rm D}-2} \sim 1/T^{{\rm D}-2} ~~~,~~~ T\sim a
\ee
in the dual pre-big-bang background, where the temperature increases
with the scale factor $a$. Going from the phase with scale factor $a$
to the dual one with scale factor $1/a$, the temperature $T$ remains
unchanged, since we applied not only a duality transformation but also
a time reversal, changing the range of cosmic time $t$ from
$[-\infty,0]$ to $[0,\infty]$.

The PBB scenario attempts to describe the very early universe near the
Planck scale, when general relativity breaks down. It should not be
considered as an alternative to the conventional standard cosmological
model supplied by inflation, but as a completion to it. Moreover, in
full string theory the pre-big-bang solution is expected to be
smoothly connected to a standard post-big-bang universe whose
evolution is described by general relativity.  One area of activity in
string cosmology studies precisely this high curvature/high coupling
phase which has to be described by string theory, either via
corrections to the low curvature/low coupling equations or by some
other means.

Another area of research deals with those observational consequences
of the PBB model which are most probably independent of this
transition and which distinguish the PBB model from other inflationary
cosmologies. One then also wants to study how well different
observables agree with present observational constraints.  The
cosmological implications of string theory, and more precisely, the
comparison of the predictions of the PBB scenario with observations,
will provide a test for string theory as a fundamental theory and fix
some of the parameters of the PBB model.  However, despite the
attractive features of the PBB scenario, one should keep in mind that
there still remain theoretical and phenomenological aspects which are
not yet fully understood. Among the phenomenological aspects of the
PBB scenario, vacuum fluctuations and the corresponding particle
production are important issues and they will be discussed in this
paper. In what follows, we calculate the spectra of the produced
particles and indicate some possible observational consequences.

We consider phenomenological aspects of the pre-big-bang solutions
derived from the low-energy string effective action. We decompose the
spatial sections of the full $D$-dimensional space-time into a
three-dimensional external expanding sub-manifold and a 
$(D-4)$-dimensional internal sub-manifold. The internal spatial
dimensions shrink down to the final compactification scale, which is
typically given by the fundamental length parameter of string theory.

At conformal time $\eta=-\eta_1$, when string corrections become
important, we make an instantaneous transition from the pre-big-bang
to the post-big-bang era where the internal dimensions as well as the
dilaton are supposed to be frozen while the three external
dimensions keep expanding like in an ordinary radiation-dominated
Friedmann universe. A crucial but reasonable assumption, is that the
spectra of the produced particles on scales much larger than $\eta_1$,
are not influenced by the details of the transition from the pre- to
the post-big-bang era. This has been verified numerically with some
toy models for the pre- to post-big-bang transition~\cite{cf}.  In
what follows, we neglect the effects of an intermediate string
phase. More precisely, in our model the universe starts at the dilaton
driven era and then enters the radiation-dominated era. It is possible
that the intermediate string phase is long and affects a considerable
region of the spectrum. Then, the spectra we calculate are still valid
for all modes which exit the horizon before the string phase.

Our paper is organized as follows.  In Section~II we discuss the
quantum creation of background perturbations (gravitons, dilatons and
moduli fields). In the context of the PBB scenario, there is an
accelerated shrinking of the event horizon during the
super-inflationary phase~\cite{pbb2}, which ends when a maximal
curvature scale is reached and higher-derivative terms can no longer
be neglected. After an intermediate string phase, the universe
enters the decelerating, radiation-dominated phase~\cite{pbb1,pbb2,sf}.
We shall see that, for generic matching conditions one obtains red
dilation and moduli 
 spectra while the graviton spectrum is blue. This is due to the fact
 that, during the pre-big-bang phase, scalar perturbations (the
 dilaton and moduli perturbations) grow
 on super horizon scales. Hence long wave-length modes which leave the
 horizon early have more time to grow. On the other hand, gravity
 waves are roughly 
 constant and given by the curvature at horizon exit which is 
 larger for the scales exiting
 later, leading to a blue graviton spectrum, in contrary to
standard inflation where curvature remains essentially constant
which implies a flat spectrum.

In Section~III we discuss the quantum creation of fields which do not
contribute to the background. We call them ``seeds''.  Seeds are
produced by the amplification of quantum fluctuations of fields which
are present in string theory but are not part of the homogeneous
background.  In particular, we study electromagnetic (EM) seeds and
their r{\^ o}le for primordial galactic magnetic fields, and
Kalb-Ramond axion  seeds and their r{\^ o}le for large-scale
structure and the cosmic microwave background (CMB) anisotropies.
Both cases are characteristic of the PBB scenario, since in the
conventional picture based on general relativity, there is no
Kalb-Ramond axion and electromagnetic perturbations cannot be excited
due to their conformal coupling to the metric and the absence of a
dilaton field.

Up to this point we assume the spatial section of the four-dimensional
homogeneous background geometry to be isotropic, in other words there
is a single scale factor. This assumption will be dropped in
Section~IV, where we analyze perturbations in four-dimensional
anisotropic pre-big-bang models. In particular, we study the creation
of axions and photons and compare the result with the isotropic
case. In Section~V we state our main conclusions.

Notation: Cosmic time is denoted by $t$, and conformal time by $\eta$;
they are related by $t=\int a d\eta$, where $a$ is the scale factor.
Correspondingly, a dot stands for derivative w.r.t. cosmic time $t$,
and a prime stands for derivative w.r.t. conformal time $\eta$.  The
scale factors are $a$ (and $b$ in the anisotropic case). The dilaton
field is $\phi$ or 
$\varphi$, the modulus field is ${\cal B}$, and the axion field is
$\sigma$.  The spectral index of the perturbation spectrum is denoted
by $n$.  The gauge-invariant Bardeen potentials are $\Phi$ and $\Psi$.  The
critical density is $\rho_{\rm c}=3M_{\rm Pl}^2H^2/(8\pi)$, where $H=
a'/a^2=\dot a/a$.  The transition scale in units of the Planck mass $M_{\rm
Pl}$ is $g_1= H_1/M_{\rm Pl}$. Since the universe is radiation
dominated at time $\eta_1$, the fraction of the total (critical) energy density
in radiation at a given time $\eta$ is
$\Omega_{\gamma}(\eta)=(H_1/H)^2(a_1/a)^4$.  Latin indices $i,j$ take
values $1,2,3$, while capital letters $A,B$ are equal to $1,...,D-4$,
where $D$ stands for the dimensionality of the space-time. The
subscripts $\rm E$, $\rm S$, $\rm A$ stand for Einstein, string and
axion frames respectively.

\section{Perturbations of the background}

In this section we study  perturbations that may be
generated by parametric amplification of vacuum fluctuations, as
the universe goes through the transition from the pre- to the
post-big-bang era. In particular, we calculate the spectrum of
quantum perturbations in the metric, dilaton and moduli fields 
produced in the classical PBB background.  We assume the 
axion field not to contribute to the background energy density; of
course quantum fluctuations of the 
axion field cannot be neglected and, as we show in the next section,
they may even lead to the observed density perturbations.  

During the pre-big-bang era, the universe goes through a phase
of accelerated evolution with growing curvature scale, and decreasing
co-moving Hubble length $|d\ln a(\eta)/d\eta|^{-1}$. Vacuum
fluctuations in a given field which are initially, $\eta\ra -\infty$,
inside the Hubble scale exit at $\eta\sim -1/k$ and enter again in the
subsequent radiation (or matter) era at $\eta\sim +1/k$.
The time evolution of the background acts like a time-dependent
potential and generically leads to particle creation.  On scales $k$
which are stretched to cosmologically large scales, this process can
lead to classical perturbations of the gravitational, dilaton and
moduli fields as in the usual inflationary scenario. The
characteristics of the generated spectrum of perturbations depend on
the evolution of the background and on the coupling of the
fields.

We consider a $D$-dimensional space-time, which contains a four-dimensional
homogeneous and isotropic external metric and a $(D-4)=m$-dimensional
compactified internal metric. It is given by 
\be 
ds^2_{\rm D}=-dt^2+g_{ij}dx^idx^j+\gamma_{AB}dx^Adx^B~, 
\ee 
where $i,j =1,2,3$, and $A,B=1,...,m=D-4$. In a four-dimensional,
spatially homogeneous
space-time the antisymmetric tensor field $H_{\mu\nu\alpha}$ has only
one degree of freedom, which can be expressed by a pseudo-scalar axion
field $\sigma$.  This ``universal axion of string theory'', is the
four-dimensional dual of the Kalb-Ramond antisymmetric tensor field
present in the low-energy string effective action~\cite{copeland}. In
the four-dimensional external space-time, the effective dilaton and
the antisymmetric tensor field are respectively~\cite{copeland}
\bea
\varphi&\equiv& \phi-m{\cal B}\nonumber\\
H^{abc}&\equiv&e^{\varphi}\epsilon^{abcd}\nabla_d\sigma~,
\eea
where the modulus field $\exp(m{\cal B})$ determines the volume of the
internal compactified space. Assuming the external four-dimensional
space-time to be described by a
Friedmann--Lema\^{\i}tre--Robertson--Walker (FLRW) metric whose
space-like sections are flat and taking, for simplicity, also for the
internal space an isotropic, homogeneous, flat, cylindrical ansatz,
$\ga_{AB}=\exp(2{\cal B})\de_{AB}$, the dilaton-moduli-vacuum solutions
(time-dependent dilaton and moduli fields, but constant axion field)
are given in terms of conformal time $\eta$ in the string frame
by~\cite{copeland}
\bea 
e^{\varphi}&\sim ~|\eta|^{\rm r} & \sim
     ~|\eta|^{(3\al -1) \over (1-\al)}  \nonumber\\ 
a&\sim ~ |\eta|^{(1+{\rm  r})/2}  & \sim
   ~|\eta|^{\al\over (1-\al)}  \nonumber\\ 
e^{\cal B}&\sim ~|\eta|^{\rm s}   & \sim
	 ~|\eta|^{\beta\over 1-\al}	~,  \label{bg}
\eea
with the Kasner constraint equation\cite{copeland} 
\be
{\rm r}^2 +2m{\rm s}^2=3 ~,~~~~~~ 3\al^2 +m\beta^2=1~.
\ee
The expressions in terms of ${\rm r,s}$ are equivalent to those in terms of
$\al,\beta$ and both can be found in the literature.
Note that $D=4$ corresponds to $r_{\pm}=\pm \sqrt{3}$. The
interesting solution describing an expanding external space is the
one with $r<0$ for which $1+r <0$ (or equivalently $\al<0$) since
$|\eta|$ is decreasing.

To study metric, scalar-dilaton and moduli fields perturbations, we
define the external pump field $P$ responsible for their amplification.
For this, we identify for each perturbation the canonical variable
$\psi$, which diagonalizes the perturbed action expanded up to second
order~\cite{pertth}. By varying the perturbed action, we find that the
Fourier modes $\psi_k(\eta)$ of each  perturbation satisfy a
decoupled, linear equation of the type
\be
\psi_k''+\left(k^2-{P''\over P}\right)\psi_k=0~.
\label{pump}
\ee
In the ``sub-horizon regime'' defined by $k^2P''/P\gg 1$, the solutions to
Eq.~(\ref{pump}) are simple plane-wave solutions; whereas in the
``super-horizon regime'' defined by $|k^2P''/P|\ll 1$, the general solution
is given by
\be
 \psi_k(\eta) \simeq A_kP(\eta) + B_k{1\over P(\eta)} 
 \int^\eta{d\tilde\eta\over P(\tilde\eta)^2} ~,~~~  |k^2P''/P|\ll 1~.
\ee
At the beginning of the pre-big-bang inflationary phase,
each perturbation is well inside the horizon,
$|k^2P''/P|\gg1$. We normalize  the plane-wave solutions of
Eq.~(\ref{pump}) in this regime to the vacuum fluctuation spectrum. 

If $P$ obeys a power-law,  $P\propto (-\eta)^p$, the correctly
normalized solution of Eq.~(\ref{pump}) during the pre-big-bang phase
can be written in terms of a Hankel function of second kind,
\be
\psi_k = \eta^{1/2}H_{\mu}^{(2)}(|k\eta|)~~, \mbox{ with }
~~\mu=\left|p-{1\over 2}\right|~~,~~\eta\leq -\eta_1~.
\ee
In the radiation-dominated era, we will find that the term $P''/P$
vanishes, implying that the solutions of Eq.~(\ref{pump}) are
free plane-wave solutions
\be
\psi_k ={1\over {\sqrt k}}\left[c_+(k)e^{-ik\eta}+c_-(k)e^{ik\eta}\right]
~~,~~\eta\geq -\eta_1.
\ee
The coefficients $c_+,c_-$ are the  Bogoliubov coefficients.
Assuming that the universe goes from the pre-big-bang phase to the
radiation-dominated era at the transition time $\eta=-\eta_1$,
matching the pre-big-bang solution at $\eta=-\eta_1$ to  the radiation
era solution at $\eta=\eta_1$ determines the  Bogoliubov coefficients.

The spectral energy density of produced particles, let us call them
$x$, is related to the coefficient $c_-(k)$ by (see~\cite{pertth})
\be
\rho_x(\omega)={d\rho_x\over d\log\omega}
 =  {\omega^4\over \pi^2}|c_-(\omega)|^2~.  \label{rhox}
\ee

The power spectrum $\De$ for energy density perturbations $\de x$ of a
given field $x$ is defined by
\be
\De_{\delta x}\equiv {k^3\over 2\pi^2}|\delta x|^2~
\ee
and the spectral index $n$ of the perturbation spectrum is defined by
\be
n-1\equiv {d\ln \De_{\delta x}\over d\ln k}~.
\ee
The value $n=1$ at second horizon crossing during the post-big-bang
era characterizes a Harrison-Zel'dovich or scale-invariant spectrum~\cite{HZ}.

We study metric, dilaton and moduli fields perturbations in the 
Einstein frame which is conformally related to the string frame by
\be
g_{ab}^{\rm E}=e^{-\varphi}g_{ab}^{\rm S}~.
\ee
While strings are minimally coupled in the string frame, both
dilaton and moduli fields are minimally coupled in the Einstein frame.

 The scale  factor in the Einstein frame, $a_{\rm E}=e^{-\varphi/2}a$,
satisfies the Friedmann equation
\be
\left({a'_{\rm E}\over a_{\rm E}}\right)^2 = {8\pi G\over
3}\left({1\over 4}\varphi'^2 + {m\over 2}{\cal B}'^2\right)~.
\ee
The fields $\varphi$ and ${\cal B}$ behave like ordinary free, massless scalar
fields in this background and satisfy the equations of motion~\cite{KS}
\bea
\varphi'' + 2{a'_{\rm E}\over a_{\rm E}}\varphi' & = & 0 \\
{\cal B}'' + 2{a'_{\rm E}\over a_{\rm E}}{\cal B}' & = & 0~.
\eea
The background in the Einstein frame is contracting like $a_{\rm E} =
\sqrt{|\eta/\eta_1|}$. 

In the Einstein frame,  first-order scalar and tensor perturbations of
the metric are given, in the longitudinal gauge, by~\cite{KS}
\be 
d s_{\rm E}^2=  a_{\rm E}^2(\eta)\left[-(1+2 \Psi)
d\eta^2 +\left
\{(1+2 \Phi)\delta_{ij}+h_{ij}\right \} dx^idx^j \right]~, 
\ee
where $\Psi,\Phi$ are the gauge-invariant Bardeen potentials,
$h_{ij}$ denote the tensor perturbations, and $\eta$ denotes
conformal time.

Metric perturbations are decomposed into scalar, vector and tensor
modes, which to first order evolve independently. We disregard vector
perturbations of the metric which, in the absence of seeds decay
quickly during the radiation-dominated era. Moreover, in the spatially
flat gauge, the evolution equations for the dilaton and moduli fields
are, to first order, decoupled from the metric perturbations.

\subsection{Scalar perturbations}

The off-diagonal components of the perturbed Einstein's equations in
the longitudinal gauge lead to the condition $\Phi=-\Psi$. The
other components imply the decoupled equation~\cite{pertth} 
\be
\Psi_k'' + 6 {a'_{\rm E}\over a_{\rm E}}\Psi_k' +k^2\Psi_k=0~,
\ee
for the Fourier mode $\Psi_k$, as well as the constraint
equation
\be
\Psi_k'+{a'_{\rm E}\over a_{\rm E}}\Psi_k=~2\pi G\varphi '(\delta\varphi)
+ 4\pi Gm{\cal B}'(\delta{\cal B}),
\ee
which relates the scalar metric perturbations to the dilaton and moduli
field perturbations, $(\delta\varphi)$ and $(\delta{\cal B})$ respectively.

Analyzing the second order perturbed action, one finds that the
variable $v_k$ defined by
\be
 v_k = a\left[{1\over 2}(\de\varphi)\varphi'+m(\de{\cal B}){\cal B}' + 
	{\varphi'^2/2+m{\cal B}'^2\over a'_{\rm E}/a_{\rm E}}\Psi_k\right]
	{1\over \sqrt{\varphi'^2/2+m{\cal B}'^2}}
\ee
is a canonical variable satisfying the equation of motion
\be
 v_k'' +(k^2- {a''_{\rm E}\over a_{\rm E}})v_k = 0~. \label{vpp}
\ee
The variable $v_k$ is related to the Bardeen potential $\Psi_k$ by
\be
  \Psi_k = {-4\pi G \sqrt{\varphi'^2/2+m{\cal B}'^2} \over k^2} 
                                      \left({v_k\over a}\right)' ~.
\ee
The correctly vacuum normalized solution of Eq.~(\ref{vpp}) is
\[ v_k(\eta) = \eta^{1/2}H_0^{(2)}(k\eta) ~. \]
This leads to the following spectrum for the Bardeen
potential~\cite{metricpert} 
\be k^{3/2}|\Psi_k|\simeq
{H_1\over M_{\rm Pl}}{|k\eta_1|^{3/2} \over |k\eta|^2} ~,  
\ee 
where $H_1\equiv H(\eta_1)$.  The spectrum of the Bardeen potential
at the end of the pre-big-bang phase is therefore red. At
$\eta=\eta_1$ we have
\be k^{3}|\Psi_k|^2(\eta_1)\simeq
g_1^2{1 \over k\eta_1} \propto k^{n_{\Psi}-1}  ~,
\ee 
implying a spectral index $n_{\Psi}=0$.
Here $g_1=H_1/M_{\rm Pl}$ is the string coupling at the end of the
pre-big-bang phase which is of the order of $10^{-3} \le g_1 \le 10^{-1}$. 
From the above
equation one might doubt whether linear perturbation theory is still valid
when $\eta \rightarrow \eta_1$ since then the Bardeen potential can
become very large. But one has to keep in mind that the Bardeen
potential is not a measurable quantity and if one analyzes physical
perturbation amplitudes like for example
$$(C_{\mu\nu\al\beta}C^{\mu\nu\al\beta})/(R_{\mu\nu\al\beta}R^{\mu\nu\al\beta})
\simeq |(k\eta)^2\Psi|^2k^3 $$
where $C$ and $R$ denote the Weyl and Riemann curvatures
respectively, one finds that the perturbation amplitudes remain small
during the pre-big- bang as long as $H_1<M_{Pl}$. 
The above quantity is also of the order of the spectral 
energy density distribution associated with the Bardeen potential.

The Bardeen potential above is generated by some cosmic energy density
perturbation satisfying the Poisson equation
\be 4\pi Ga^2\rho_{x} \de_x(k) = k^2\Psi_k ~, \label{Poiss}\ee 
and thus leading to a spectral distribution
\be k^{3}|\de_x(k)|^2 \simeq  g_1^2(k\eta_1)^3 ~. \ee
Since $|k\eta_1| < 1 $, for all amplified frequencies, $\de_x(k)$ is
always a small contribution to the total energy density as long as 
 $g_1 < 1$.
A more detailed study of scalar metric perturbations can be found
in Brustein et al. (1995b), where the matching of pre- to
post-big-bang perturbations is discussed. There it is argued that this
dominant term of the Bardeen potential is converted entirely into the
decaying mode at the pre- post-big-bang transition and that only the
subdominant term with a blue, $n=4$ spectrum survives. Recently
(Durrer and Vernizzi, 2002) it has been shown that this result is true
only for very specific matching conditions, that however, in the
generic case, the naively expected dominant contribution with a red
spectrum, $n_\Psi =0$ is induced in the radiation era.

Before discussion the implication of this finding, let us derive it by a
different but equivalent approach. We study scalar
perturbations in this background by direct quantization of the
dilaton and moduli fields which contribute the energy density
perturbation $\de_x$. By consistency, this leads to the same
spectral index, as we will now compute. 
We again work in the Einstein frame, in which both the
dilaton and moduli fields evolve as  minimally coupled massless
fields. In the spatially flat gauge, the dilaton perturbations are
decoupled from the axion perturbations and the evolution equation is
Eq.~(\ref{pump}) with 
\be 
\psi_k=a_{\rm E}(\delta\varphi)_k~~~~,~~~P=a_{\rm E}~,~~ \mbox{ hence
    }~~ p=1/2~.
\ee 
The moduli perturbations
respectively obey, for each Fourier mode, the simple wave equation,
Eq.~(\ref{pump}), with 
\be 
\psi_k={\sqrt m}a_{\rm E}(\delta{\cal B})_k~~~~,~~~P=a_{\rm E}~;
\ee 
$m$ stands for the number of internal compact dimensions.

Using  $a_E \propto |\eta|^{1/2}$, the general solution of the evolution
equation normalized to a vacuum fluctuation spectrum at $\eta\ra
-\infty$, can be written 
in terms of the zeroth Hankel function of the second kind as 
\be
\psi_k=\eta^{1/2}H_0^{(2)}(|k\eta|)~~,~~\eta\leq -\eta_1~.  \label{H0}
\ee 
In the radiation era, which follows the dilaton-driven era, one has 
instead free-plane wave solutions 
\be 
\psi_k={1\over {\sqrt k}}\left[c_+(k)e^{-ik\eta}+c_-(k)e^{ik\eta}\right]~.  
\label{pw}
\ee
The density parameter of produced dilatons per logarithmic frequency 
interval is determined by matching the solution~(\ref{H0}) to (\ref{pw})
and using Eq.~(\ref{rhox}) (see also Gasperini and Veneziano, 1994):
\be
\Omega_\varphi(\omega,\eta)={\rho(\omega)\over \rho_{\rm c}}
={1\over \rho_{\rm c}}{d\rho_\varphi\over d\log\omega}
\simeq g_1^2\left(\omega\over \omega_1\right)^3
\left(H_1\over H\right)^2\left(a_1\over a\right)^4 =
g_1^2\left(\omega\over \omega_1\right)^3\Om_\ga ~, \label{Omdm}
\ee
where $\rho(\omega)$ denotes the spectral energy density of
produced dilatons, and $\omega_1=k_1/a_1=1/(a_1|\eta_1|)$
represents the maximal amplified frequency. 

More precisely, on super horizon scales, the power spectrum of dilaton
and moduli perturbations is (see Copeland et al., 1997a) 
\bea
\De_{\delta\varphi}
&\sim & {2\over \pi^3}H^2
(k\eta)^3~[\ln(k\eta)]^2\nonumber\\
\De_{\delta {\cal B}}
&\sim & {2\over m\pi^3}H^2
(k\eta)^3~[\ln(k\eta)]^2~,
\eea
respectively. Thus, the amplitude of dilaton and moduli perturbations
grows towards small scales.

Both the dilaton and moduli fields  have steep blue energy perturbation 
spectra~(Copeland et al., 1997a),
\be
n_{\varphi}=n_{\cal B}=4~.
\ee
According to Eq.~(\ref{Omdm}), the contribution of these perturbations
to the background density at the upper cutoff $\om=\om_1$ is of the
order $\sim g_1^2\ll 1$, on larger scales (lower frequencies) this
contribution decreases like $(\om/\om_1)^3$ and hence it is completely
negligible on all scales of cosmological interest. (Note that the scale
$\la_1=2\pi/\om_1$ has expanded to about $\la_1(\eta_0)\sim 0.01$cm
today.) However, the induced perturbation in the geometry is related to
$\Delta_{\de\varphi}$ and $\De_{\de{\cal B}}$ via the Poisson equation
(\ref{Poiss}) and therefore has a red spectrum, $n_\Psi=n_{\varphi}-4
= n_{\cal B}- 4 =0 $.

The relevant question is now which of these spectra is inherited in
the subsequent radiation era. This question can only be
answered by studying the matching condition as it has been done in
Durrer and Vernizzi (2002). There it was found, that generically the
Bardeen potential in the radiation era is roughly matched to the one
in the pre-big-bang era and hence the relevant spectral index is
$n=n_\Psi=0$. Only if the matching takes place at the constant energy
surface, and if the effective surface tension needed to convert
contraction into expansion is unperturbed, can the dominant term be
converted entirely into the decaying mode. In this case one obtains
$n=n_{\varphi}=n_{\cal B}=4$. 

If the red spectrum is inherited in the radiation era, curvature
perturbations soon become very large, at horizon crossing: during the
radiation era one has
$$\left. (C_{\mu\nu\al\beta}C^{\mu\nu\al\beta})/(R_{\mu\nu\al\beta}
  R^{\mu\nu\al\beta})\right|_{k\eta\sim 1}
\simeq |\Psi|^2k^3 \simeq g_1^2{1 \over k\eta_1} \gg 1$$
for cosmologically relevant scales. Perturbations become large, in
conflict with observations!

In the contrary, if the dominant mode is entirely converted into the decaying
mode of $\Psi$, we have 
$$\left. (C_{\mu\nu\al\beta}C^{\mu\nu\al\beta})/(R_{\mu\nu\al\beta}
R^{\mu\nu\al\beta})\right|_{k\eta\sim 1}
\simeq |\Psi|^2k^3 \simeq g_1^2(k\eta_1)^3 \ll 1$$
for cosmologically relevant scales. In this case 
it is unlikely that the dilaton and moduli quantum
fluctuations from the pre-big-bang have left any characteristic
observable signature in the present universe.

It has been shown by Durrer and Vernizzi (2002) that dilaton (or
moduli) perturbation spectra actually can become scale invariant if an
exponential  potential for the dilaton (or the moduli), 
$V=-V_0\exp(-\la\varphi)$ is added to the
Lagrangian. Again, if matching takes place on the constant energy
hypersurface, a blue spectrum, in this case $n=3$ is inherited;
otherwise one finds $n=1$. This modified pre-big-bang model therefore
can produce scale-invariant adiabatic fluctuations during the radiation era.

\subsection{Tensor metric perturbations}

For each Fourier mode, the evolution of tensor metric perturbations in the 
Einstein frame satisfies to lowest order Eq.~(\ref{pump}) with 
\be
\psi_k={1\over \la_{\rm s}}a_{\rm E}(\delta h)_k~~~~,~~~P=a_{\rm E}~,
\ee
where $\lambda_{\rm s}=\sqrt{\alpha' \hbar}$ denotes the 
short-distance cut-off of string theory and $a_E \propto |\eta|^{1/2}$
is the well-known expansion law for a scalar field
dominated Friedmann universe, independent of the expansion/contraction
of the internal dimensions. We note that in the
S-frame, the string length parameter $\lambda_{\rm s}$ is constant.
We have to divide $\delta h$ by  $\lambda_{\rm s}$ for
dimensional reasons, the canonical field must have the dimensions of a
scalar field. Like always, one finds $\psi_k$ by writing the second
order perturbation of the action in canonical form~\cite{grwa}.
The solutions of the evolution equation are as in the last sub-section
 \bea
\psi_k&=&\sqrt{|\eta|}H_0^{(2)}(k\eta)~~ \mbox{ in the pre-big-bang era},\\
 \psi_k&=&{1\over
{\sqrt k}}
\left[c_+(k)e^{-ik\eta}+c_-(k)e^{ik\eta}\right]~~,\mbox{ in the
  post-big-bang era}
\label{tmp}
\eea 
Matching these solutions at the onset of the radiation-dominated era,
$\eta=\eta_1$  determines the Bogoliubov coefficients.
The magnitude of $c_-$ gives the amplification 
of the gravitational waves with respect to the minimal vacuum fluctuation. 

The spectrum depends only on the dynamics of the scale factor in the
Einstein frame, which can be parameterized generically 
as $a_{\rm E}(\eta)=(-\eta)^{\ga}$. The case of a dilaton and moduli
background corresponds to $\ga=1/2$. If a dilaton potential is added,
we have $0<\ga<1/2$. 
For $\ga \le 1/2$, the co-moving amplitude $(\de h)_k$ approaches a constant
asymptotically ($|k\eta|<<1$)~\cite{grwa}. The perturbation amplitude
$\delta_h(k)$ can be expressed in terms of the Hubble constant at
horizon crossing (HC), defined by $|k\eta|\sim 1$, as in~\cite{pertth}
\be 
|(\delta h)_k|\equiv k^{3/2}|h_k| \sim 
\left({H\over M_{\rm Pl}}\right)_{\rm HC} \sim   g_1(k\eta_1)^{3/2}~.
\ee 
The amplitude remains constant in time. However, since
higher-frequency modes cross the horizon at later times, therefore at a
higher value of $H$ if the curvature scale is growing, their amplitude
is enhanced with respect to lower-frequency modes. This is 
different from the usual case of a scale-invariant spectrum, where
the amplitude remains the same for all modes.

The amplitude of tensor perturbations over scales $k^{-1}$ 
varies in time according to
\be
|(\delta h)_k(\eta)|\sim g_1
(k\eta_1)^{3/2} \ln|k\eta|~.  
\ee
A robust prediction of the PBB scenario is the finding that the
spectrum of gravitational waves is characterized by rising amplitude
with increasing frequency. The spectrum of primordial gravitational waves
is steeply growing on short scales with a spectral
index $n_{\rm T}=3$, in contrast to the more conventional inflation
models, for which $n_{\rm T}\lsim 0$.
  If the duration of the intermediate string scale can be neglected,
this steep blue spectrum leads to gravitational wave amplitudes of the
order $|(\delta h)_0(f)|^2\sim g_1^2(10^{11}Hz/f)^3$, far beyond the
detection limits of any gravitational wave experiment in
consideration. However, if the intermediate string scale is
sufficiently long, the form and the amplitude of the tensor
perturbation spectrum can
be modified (flattened) on small scales, and thus this conclusion can be
altered (for a detailed discussion see~\cite{Mag}).

\section{Perturbations of fields which do not contribute to the
         background: Seeds}
In the previous section we have discussed the perturbations obtained
in the  components which are also present in the background,
the dilaton, the moduli and the gravitational field. Here we study
perturbations of the fields which do not contribute to the background.

 All quantum fields, even if their expectation value vanishes, exhibit
quantum fluctuations. During an inflationary era, they  stretch
beyond the horizon scale and ``freeze in'' as classical non-vanishing
inhomogeneous field configurations. Their contribution is in general
second order (or higher) in the  field perturbation, but it can
nevertheless lead to appreciable perturbations in space-time.

Such second order perturbations can induce cosmic structure formation
after the transition from the pre-big-bang to the radiation-dominated
era in two ways. Either they decay into dark matter particles and
just leave their ``imprint'' as the initial condition for the dark
matter and radiation perturbations. A model of this kind has been
studied in Refs.~\cite{pe1,pe2}. Its main difference to ordinary cold
dark matter (CDM)-like models is that the resulting perturbations are
of isocurvature nature and that they do not obey Gaussian
statistics.  Or, the perturbation decouple from CDM
and radiation and remain active as a ``seed'' which induces
geometrical perturbations and perturbations in the radiation and CDM
components, solely by gravitational interaction~\cite{d90}. This is
the possibility which has been studied in the case of axion field
perturbations in Refs.~\cite{1,2,afrg,fra}.

As we shall point out in the next sub-section, second order
perturbations can also remain in the form of large-scale coherent
magnetic fields and may contribute to the resolution of the problem of
the origin of primordial magnetic fields~\cite{kronberg}.

\subsection{Electromagnetic seeds and primordial magnetic fields}
Let us first consider the photon field in the pre-big-bang universe.
Since photons are conformally invariant they do not couple to a
homogeneous and isotropic metric but they couple to the
dilaton. Therefore, in contrast to ordinary inflation there is photon
production in pre-big-bang inflation.

The ``pump field'' for photon perturbations is $P^{EM}=e^{-\varphi/2}$ with
$\varphi=-2\ga\log(-\eta/\eta_1)$,  according to Eq.~(\ref{bg}),
$\ga=-(3\al-1)/2(1-\al)$. Setting $\psi= e^{-\varphi/2}A_\mu$, the 
photon equation of motion in the radiation gauge, $A_0=0$ and $\dd^iA_i
=0$ in $k$-space reduces to
\be
\psi''_k + \left[k^2- {\ga(\ga-1)\over \eta^2}\right]\psi_k=0.
\label{Elm}
\ee
The general solution of this equation is a linear combination of the
Hankel functions $\eta^{1/2} H^{(1)}_\mu(k\eta)$ and  $\eta^{1/2}
H^{(2)}_\mu(k\eta)$  with $\mu=|\ga-1/2|$. 
At very early times, $\eta\ll -1/k$, the field is supposed to
be in the in-coming vacuum state, 
\[ \psi_k =  \sqrt{{2\over \pi k}} e^{-ik\eta} ~.\]
Therefore, the correctly normalized initial  vacuum
fluctuation spectrum, can be written in terms of the Hankel function
of the second kind,
\be
\psi_k= |\eta|^{1/2} H^{(2)}_\mu (k\eta) ~, \,\,\,\,\,\,\,\,\,\,
\eta < -\eta_1 ~.
\label{210}
\ee
It describes an in-coming vacuum which, as the perturbation becomes
super horizon ($-k\eta <1$), becomes a non-trivial classical field
configuration.

In the radiation era the dilaton is constant and $\psi$ obeys an
ordinary wave equation with solution
\be
\psi_k= {1\over \sqrt k}\left[c_+(k) e^{-ik\eta}+
c_-(k) e^{ik\eta}\right] ~, \,\,\,\,\,\,\,\,\,\,\,\,
\eta >\eta_1 
\label{211}
\ee
Using Eq.~(\ref{210}) as initial condition on
super-horizon scales, $k\eta\ll 1$, and for $\eta\gg \eta_1$, we find
\be
c_\pm=\pm c(k) e^{\pm ik\eta}~, ~~~~~~~~~
\psi_k= {c(k)\over \sqrt k}\sin k(\eta-\eta_1)~, ~~ ~~~~~~~
|c(k)|\simeq (k/k_1)^{-\mu-1/2} ~,
\label{212}
\ee
 where $k_1=1/|\eta_1|$ represents the maximal amplified
frequency (higher-frequency modes are not amplified).
According to Eq.~(\ref{rhox}) the associated energy-density distribution
 of the produced photons is 
\be
{d\rho(k)\over d\log k}\simeq \left(k\over a\right)^4
|c_-(k)|^2 \simeq \left(k_1\over a\right)^4
\left(k\over k_1\right)^{3-2\mu} ~, \,\,\,\,\,\,\, k<k_1~,~~~~~\mu<3/2~,
\label{213}
\ee
where we require $\mu <3/2$ (which is always satisfied for
$-\sqrt{3}\le r\le 0$) to avoid photon overproduction which would
destroy the homogeneity of the classical background. The
amplitude $c(k)$ has been estimated modulo numerical factors of
order 1. At large times, $\eta \gg |\eta_1|$, we thus obtain in string
cosmology,  a cosmic background of electromagnetic fluctuations which
is always blue, $n\ge 1.5$. This is too rapidly decaying to be of
cosmological relevance {\em e.g.} as seeds for the observed large
scale magnetic fields (see next section). 

This conclusion is modified
if there is a long intermediate string phase.  During such a string
phase, photon seeds are characterized by a rather flat spectrum, $\ga~
\le ~2$, and could provide the long-sought  origin of the galactic
magnetic fields \cite{7}.  

The amplified
fluctuations acquire stochastic correlation functions as a
consequence of their quantum origin.

In the radiation era we therefore have the following spectrum of
electromagnetic perturbations
\be
A_i (\bk, \eta) = {c_i (\bk)\over \sqrt k} \sin k\eta ~,
\,\,\,\,\,\,\,\,\, k_iA_i=0~, \,\,\,\,\,\,\,\,\, A_0=0 \,.
\label{21}
\ee
$A_i$ is a Gaussian random variable which obeys the stochastic
average condition:
\be
\langle A_i(\bk)A_j^\ast(\bk')\rangle= {(2\pi)^3\over 2}
\de^3(k-k')\left(\de_{ij}-{k_ik_j\over k^2}\right)
\left|{\bf A}(\bk, \eta)\right|^2~.
\label{22}
\ee
The above condition has been normalized in such a way that
\be
\sum_i\langle A_i(\bk)A_i^\ast(\bk')\rangle= {(2\pi)^3}
\de^3(k-k')\left|{\bf A}(\bk, \eta)\right|^2 ~.
\label{23}
\ee
Taking into account that the electric component of the stochastic
background is rapidly dissipated due to the high conductivity of the
cosmic plasma \cite{10}, only the  magnetic field survives. Setting
$B_i(k)= i\ep_{ijl}k_jA_l(k)$, the
condition (\ref{22}) implies
\be
\langle B_i(\bk)B_j^\ast(\bk')\rangle= {(2\pi)^3\over 2}
\de^3(k-k')\left(\de_{ij}-{k_ik_j\over k^2}\right)b^2(k,\eta) ~,
\label{24}
\ee
where
\be
 b^2(\bk, \eta) =
k^2\left|{\bf A}(\bk, \eta)\right|^2 (a_1/a)^4=
k\left|{\bf c}(\bk)\right|^2 \sin^2 k\eta ~.
\label{25}
\ee
Here we have used that, within the magneto-hydrodynamic limit, magnetic
fields in a Friedmann universe just conserve the flux per unit area and
thus scale like $1/a^2$ (see, e.g. Ref.~\cite{BEO}).

In a process of photon production, the coefficient
$\left|{\bf c}(\bk)\right|^2$
represents the Bogoliubov coefficient \cite{pertth} fixing the  average
photon number density, $\langle n(k) \rangle $, and is linked to the
spectral energy distribution by
\be
{d\rho(k)\over d\log k}= \left(k\over a\right)^4
{\langle n(k) \rangle\over \pi^2} \simeq
\left(k\over a\right)^4
{\left|{\bf c}(\bk)\right|^2\over \pi^2}~.
\label{26}
\ee
The spectrum  $\left|{\bf c}(\bk)\right|^2$ is given by
\be
\left|{\bf c}(\bk)\right|^2=\left\{\begin{array}{ll}
	 \left(k/k_1\right)^{-2\mu-1} ~,& k\le k_1, ~~\mu\le 3/2\\
	0 ~,& k> k_1 ~.\end{array} \right.
\label{28}
\ee
At first sight one might think that for $\mu=3/2$ this also induces a
scale-invariant (Harrison-Zel'dovich) spectrum of metric
perturbations, but as it has been shown in Ref.~\cite{1} this is not
the case. The reason is precisely the conformal coupling of photons.

\begin{figure}[ht]
\centerline{\epsfxsize=2.2in  \epsfbox{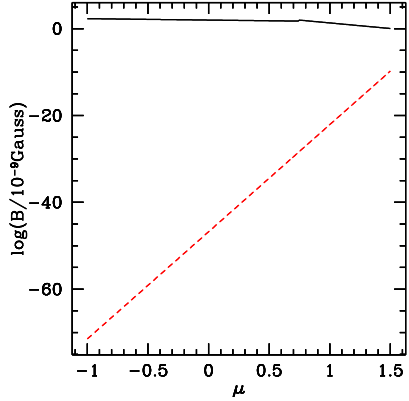}}
\caption{\label{fig1}
The magnetic field (in units of $10^{-9}$Gauss) induced in pre-big-bang 
cosmology is compared 
with the limit from CMB anisotropies. For all admitted values
of the spectral index $\mu$ the induced field (solid line) is well
below the limit (dashed line).}
\end{figure}

The result (\ref{25}) can be used to constrain $(k_1,\ga)$. From the
CMB anisotropies induced by the gravitational coupling of the magnetic
field one obtains~\cite{DFK}
\[  B_1 <7.9\times10^{-6}e^{3n}{\rm Gauss}~,\]
where $B_1$ is the magnetic field amplitude at scale $\la=0.1h^{-1}$Mpc and $n$
is the spectral index of the magnetic field spectrum. In our case 
$n=-2\mu$ and (see Ref.~\cite{DFK})
\bean B_1^2 &\simeq& {4\over (2\pi)^5}\la^{-4}(k_1\la)^{2\mu+1} \\
\log(B_1/(10^{-9}{\rm Gauss})) &\simeq& 41.2 +24.5\mu ~,
\eean
where we have used  $\la=0.1h^{-1}$Mpc and $k_1=10^{18}$GeV$/z_1$ with
$z_1 = T_1/T_0 = 10^{18}$GeV$/(2.5\times 10^{-4}$eV$) = 4\times 10^{30}$
for the last equality (see also Fig.~\ref{fig1}).

\subsection{Axion seeds, large-scale structure and CMB anisotropies}
The Kalb-Ramond (KR) axion evolves as a massless minimally coupled
field in the axion (A) frame, related to the Einstein frame by the
conformal transformation
\be
g_{\alpha\beta}^{\rm A}=e^{-2\varphi}g_{\alpha\beta}^{\rm E}~.
\ee
We will use the axion frame to study Kalb-Ramond axion perturbations.
Defining
\be \psi_A = ae^{\varphi/2}\si \equiv a_A\si~,\ee
 we are led to the canonical equation
\be
 \psi''_k+\left(k^2-{a''_A\over
a_A}\right) \psi_k =0~,
\label{42a}
\ee
very similar to Eq. (\ref{Elm}). The same procedure as in the
electromagnetic case then leads to the spectrum (\ref{213}) with
$\mu=|r_A|$, where $r_A$ parameterizes the three-dimensional axion scale
factor as $a_A(\eta)\sim \eta^{r_A+1/2}$.
For $r_A=-3/2$, in particular, the axion metric describes a de
Sitter inflationary expansion, and the energy density of a
massless KR axion background has a flat spectral distribution,
$d\rho/d\log k\simeq (k_1/a)^4$, as first noted in Ref.~\cite{8}.
The value of $r_A$ depends on the number and on the kinematics of the
internal dimensions, and the value $-3/2$ can be obtained, in
particular, for a ten-dimensional background with symmetric expansion
and contraction, $\al=-1/3$ and $\beta=1/3$ \cite{Buon}. 
In the axion case, however, the low frequency tail of the spectrum is
further affected by the radiation $\ra$ matter transition, as the axion
pump field $a_A$ is not a constant (unlike the dilaton) in the
matter-dominated era, where $a_A=a\propto \eta^2$.

 In the radiation era, i.e. for $\eta_1<\eta<\eta_{eq}$, the
effective potential $ a''_A/a_A$ is vanishing, as $\varphi=const.$ and
$a_A=a\sim \eta$, and $\psi$ is given by the plane-wave solution
(\ref{211}).
In the final matter-dominated era, i.e. for
$\eta>\eta_{eq}$, we have $a\sim \eta^2$, and 
$a''_A/a_A=2/\eta^2$. The plane-wave solution is still valid for modes
with $k>k_{eq}=\eta_{eq}^{-1}$, which are unaffected by the last
transitions. Modes with $k<k_{eq}$ feel instead the effect of the
potential in the matter era, and the general solution of Eq.~(\ref{42a}), 
for those modes, can be written as
\be
\psi_k(\eta)={\sqrt{k\eta}\over \sqrt k}\left(AH_{3/2}^{(2)}+B
H_{3/2}^{(1)}\right)
~~~~~~k<k_{eq}~, ~~~~\eta>\eta_{eq}~.
\label{337}
\ee
The matching of the solutions at $\eta_1$ determines the coefficients
$c_\pm(k)$ as in Eq.~(\ref{212}). The matching at $\eta_{eq}$ gives
\be
A+B \sim c({\bf k})\left(k\eta_{eq}\right)^{-1}~~ ,~~~~~~~~~
A-B \sim c({\bf k})\left(k\eta_{eq}\right)^{2}~~,
\label{338}
\ee
In the matter-dominated era, i.e. for $\eta>\eta_{eq}$, we
can then approximate the produced stochastic axion background as
follows:
\bea
\si (\bk, \eta) &\simeq& {c (\bk)\over a\sqrt k} \sin k\eta ~~,
~~~~~~~~~~~~~~~~k>k_{eq}~, \nonumber \\
&\simeq&{c (\bk)\over a\sqrt k}\left(k\over k_{eq}\right)^{-1}(k\eta)^2~,
~~~~~k<k_{eq}~~, ~~~k\eta<1~, \nonumber \\
&\simeq&{c (\bk)\over a\sqrt k}\left(k\over k_{eq}\right)^{-1}~,
~~~~~~~~~~~~k<k_{eq}~~, ~~~k\eta>1~.
\label{339}
\eea
The correlation functions for the various components of the stress
tensor
\be
T_\mu^\nu=\dd_\mu\si\dd^\nu\si-{1\over 2}\de_\mu^\nu
\left(\dd_\a \si\right)^2
\label{340}
\ee
of massless KR axions can be computed by exploiting the
stochastic average conditions of the Gaussian variables $\si, 
\si'$ and $\si_j=\dd_j \si$:
\bea
&&
\langle \si(\bk)\si^\ast(\bk')\rangle= {(2\pi)^3}
\de^3(k-k')\Si_1(\bk, \eta) ~,
\nonumber\\
&&\langle \si' (\bk){\si' }^{\ast} (\bk')\rangle= {(2\pi)^3}
\de^3(k-k')
\Si_2(\bk, \eta) ~,
\nonumber \\
&&\langle \si_i(\bk)\si_j^\ast(\bk')\rangle= k_ik_j{(2\pi)^3}
\de^3(k-k')
\Si_1(\bk, \eta) ~,
\nonumber \\
&&\langle \si_j(\bk){\si' }^\ast(\bk')\rangle=
-\langle \si' (\bk)\si_j^\ast(\bk')\rangle=
{i} k_j{(2\pi)^3}\de^3(k-k')
\Si_3(\bk, \eta)~;
\label{341}
\eea
the explicit form of $\Si_1, \Si_2, \Si_3$ can be found in Ref.~\cite{1}.
For example for the axion energy density we find
\be
\rho_\si={1\over 2 a^2}\left[\dot{\si }^2 +({\dd}_i
\si)^2\right] ~,
\label{345}
\ee
which leads to the spectrum
\bea
k^3\langle|\rho_\si|^2\rangle&=&{2k^3\over (2a^2)^2}
\int_0^{k_1}{d^3p\over (2\pi)^3}\Bigg[
\Si_2({\bf p})\Si_2({\bf k}-{\bf p})
+\left|{\bf p}\cdot (\bk-{\bf p})\right|^2
\Si_1({\bf p})\Si_1({\bf k}-{\bf p})
\nonumber\\
\qquad &-&
2 {\bf p}\cdot (\bk-{\bf p})
\Si_3({\bf p})\Si_3({\bf k}-{\bf p})\Bigg]~.
\label{346}
\eea
Similarly the other components of the energy-momentum tensor can be
expressed in terms of $\Si_1, \Si_1$ and $\Si_3$ (see Ref.~\cite{fra}).
Simple approximations for the functions $\Si_i$ can be found in
Ref.~\cite{1}.

The spectrum  of $\langle|\rho_\si|^2\rangle$ on scales $k\ll
k_1$, mainly depends on the behavior of the integral in
Eq.~(\ref{346}). If this integral converges for $k=0$, then we
always obtain a white noise spectrum  
$\langle|\rho_\si|^2\rangle ={\rm const.}$; this is the case for
$\mu < 3/4$. Only if the integral diverges for $k=0$  we can get a
non-trivial spectrum (with $0<n\le1$) for
$\langle|\rho_\si|^2\rangle$. 
 
Using  linear perturbations of Einstein's equations one can 
determine the induced geometrical and dark matter perturbations.  For
the induced Bardeen potentials one can derive (after a lengthy
calculation, see Ref.~\cite{1}) the 
following approximate expression, which is valid on large scales
($k\eta\ll 1$) during the matter-dominated era ($\eta>\eta_{eq}$):
\be
k^{3/2}\left|\Psi- \Phi\right| \simeq
\left\{ \begin{array}{l}
g_1^2\Om_\ga(\eta)(\om/\om_1)^{-1/2}
(\om_1/H)^{-2}\left[1+\xi
(\om_{eq}/\om_1)^2(\om_1/H)^{2\mu+1/2}\right]~,~~ \mu\le 3/4 ~,\\
g_1^2\Om_\ga(\eta)(\om/\om_1)^{-1/2}
(\om_{eq}/\om_1)^2(\om_1/H)^{2\mu-3/2}~, ~~~  3/4\le \mu\le 3/2 ~,
	\end{array} \right.
\label{353}
\ee
where $\Omega_\gamma(\eta)=(H_1/H)^2(a_1/a)^4$ is the 
density parameter of the background radiation
in the post-big-bang era, $\om_1=k_1/a$ is the redshifted string scale and
 $\xi$ is a number of order unity~\cite{1}. Note that $H_1 \simeq
\om_1(\eta_1)$. 
The second case of Eq.~(\ref{353})  can also be written as
\be
k^{3/2}\left|\Psi- \Phi\right| \simeq \begin{array}{ll}
g_1^2\Om_\ga(\eta_0)(\eta_0/\eta_{eq})^2(k\eta)^{2\mu-7/2}
  (k/k_1)^{3-2\mu}~~, & 3/4\le \mu\le 3/2 ~.
	\end{array} 
\label{Psi}
\ee
The above approximations are valid on super-horizon scales, $k\eta<1$. On
sub-horizon scales the ``seed'' energy-momentum tensor decays quickly
due to the oscillatory behavior of the solution $\psi_k$, and the
gravitational potential remains constant. On sub-horizon scales we
therefore expect
\be
k^{3/2}\left|\Psi- \Phi\right| \simeq g_1^2\Om_\ga(\eta_0)
  (\eta_0/\eta_{eq})^2(k/k_1)^{3-2\mu}~~,~~  3/4\le \mu\le 3/2 ~.
\label{Psisub}
\ee
This spectrum is scale-invariant, $k^3\left|\Psi-
\Phi\right|^2=$const. $\propto k^{n-1}$, for $\mu=3/2$. Hence,
massless Kalb-Ramond 
axion seeds with $\mu=3/2$ can induce a Harrison-Zel'dovich spectrum
of geometrical perturbations as it has been observed by the DMR
experiment aboard the COBE satellite~\cite{COBE}.
For $\mu<3/2$ one obtains tilted spectra with $0\le n\le 1$ where
$n=0$ for $\mu\le 3/4$.
  
For massive Kalb-Ramond axions the situation is qualitatively
different if the axion mass is such that all super-horizon modes 
at the time of decoupling are already non-relativistic, $k/a <m$. In this case
the contribution to temperature anisotropies is controlled by the
axion mass, and a slightly blue spectrum is still compatible with the
experimental constraints, provided the axion mass lies inside an
appropriate ultra-light mass window with an upper limit of $10^{-17}
eV$~\cite{1}. This upper limit is rather low, thus if KR axions are
heavier than $10^{-17}eV$, then they are incompatible with the 
data. However, if one considers more complicated cosmological
backgrounds, it turns out that the bounds on the axion mass can be
relaxed~\cite{axionmass}. In particular, if one allows for an axion
spectrum which grows monotonically with frequency, but with a
frequency-dependent slope, then the non-relativistic KR axions can
have a mass up to the $100 MeV$ range~\cite{axionmass}. This can be
realized, for example,  if the accelerated pre-big-bang
evolution consists of at least two distinct eras~\cite{axionmass}.

To do a more precise numerical calculation, we have to write down the
perturbed Einstein and matter equations which are of the form
\be
 \DD_k X_k = \SS_k ~,
\ee
where $X_k$ is a long vector containing all the background perturbation
variables, like the $a_{lm}$'s of the CMB anisotropies, the dark
matter density fluctuation, the peculiar velocity potential etc., $\DD$
is a linear ordinary differential operator and $ \SS_k$ is given by
the energy-momentum tensor of the ``seed'', the axion in this case. The
generic solution of this equation is of the form
\be
X(\bk,\eta_0) = \int_{\eta_{in}}^{\eta_0}\GG(\bk,\eta_{0},\eta)
	\SS(\bk,\eta)d\eta .
\label{Green}
\ee
We want to determine power spectra or, more generally, quadratic
expectation values which are then given by
\bea
\langle X_i(\bk ,\eta_0)X_j(\bk,\eta_0)^*\rangle 
&=& \int_{\eta_{in}}^{\eta_0}\int_{\eta_{in}}^{\eta_0}
	\GG_{il}(\eta_{0},\eta)\GG_{jm}^*(\eta_{0},\eta') 
 \nonumber \\
&&	\times\langle \SS_l(\eta)\SS_m^*(\eta')\rangle d\eta d\eta' .
\label{pow}
\eea
(Sums over double indices are understood.)

We therefore have to compute the {\em unequal time correlators},
$\langle \SS_l(\eta)\SS_m^*(\eta')\rangle$, of the seed
energy-momentum tensor. This problem can, in general, be solved by an
eigenvector expansion method~\cite{Turok,DKM}.  The numerical solution
for the CMB anisotropies obtained in Ref.~\cite{fra} is shown in
Fig.~\ref{cmb} below. The spectral index is chosen to be slightly
blue, $\mu=1.425$, to fit the data  better. The cosmological parameters
chosen for the plot are $\Om_\La=0.85$, $\Om_m=0.4$, $h=0.65$ and
$\Om_bh^2=0.02$. We compare the numerically obtained spectrum with the
COBE measurement~\cite{TH} and recent CMB 
anisotropy data, Boomerang98~\cite{Boom} and Maxima~\cite{Maxi}.
\begin{figure}
\centerline{\epsfxsize=2.9in  \epsfbox{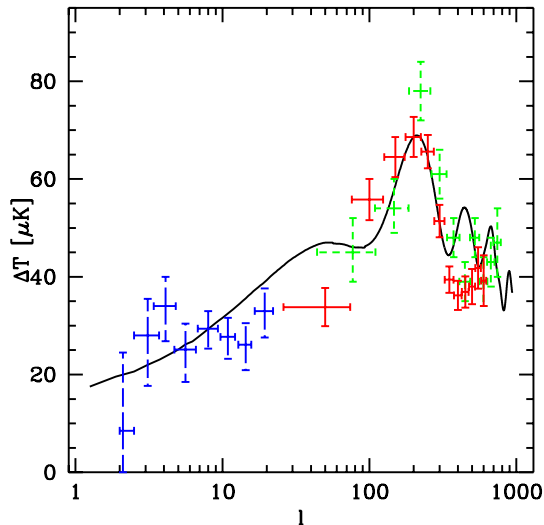}}
\caption{\label{cmb} The CMB anisotropies obtained from axionic seeds
are compared with observations. The data used are the COBE data
(dashed) on large scales, and the Boomerang98 data (solid) as well as
the Maxima-1 data (dashed) on intermediate scales.}
\end{figure}

Axionic seeds lead to isocurvature perturbations. To fit the observed
position of the first acoustic peak it therefore requires a closed
universe. The values adopted for Fig.~\ref{cmb} are also in agreement
with the recent supernovae results which indicate an accelerating
universe~\cite{PeSc}. 

We consider this an important result which is detailed in
Ref.~\cite{fra}: a slightly blue spectrum of isocurvature
perturbations in a closed universe with considerable cosmological
constant can fit relatively  well the first peak of present CMB data but are
  poor fits of the secondary peak structure. The dark matter power spectrum
is in qualitative agreement with, {\em e.g.} the APM data.

\section{Perturbations in an anisotropic pre-big-bang model}

Recently there has been some evidence that the dilaton
driven inflationary stage is generically spatially anisotropic.
In Ref.~\cite{bdv} it has been proposed that the universe starts 
as a bath of stochastic gravitational and dilatonic waves.
Some of these will be strong enough and collapse
due to self gravity forming a black hole.
A collapsing solution in the Einstein frame
corresponds to an expanding solution in the string frame.
Each of these black holes can be interpreted as a
pre-big-bang bubble pinching off the spacetime 
and creating a new universe. However, the solution
close to the singularity inside a black hole is generically
described by a Kasner metric which is spatially homogeneous over
some region but not necessarily isotropic.
In Ref.~\cite{fkvm} a different realization of this proposal was given.
Here pre-big-bang bubbles are created in the interaction region
of two colliding plane waves. In the interaction region a singularity
is generically formed which again described by an anisotropic Kasner metric.

The fate of global anisotropy in spatially homogeneous backgrounds
during the dilaton driven inflationary stage was studied in
Ref.~\cite{kd}. It was found that these persist and are {\it not}
inflated away as in the usual potential-dominated inflationary
scenarios.  Therefore it seems to be important to investigate if there
are any observational imprints from an anisotropic dilaton driven
inflationary stage.  We will consider axionic and electromagnetic
density perturbation spectra.  In order to simplify, only axi-symmetric
backgrounds will be considered. This, of course, means to impose an
additional symmetry, but on the other hand makes the whole problem
more tractable.

\subsection{Axion production}

Let us consider a four-dimensional PBB cosmological model
with the line element
\begin{eqnarray}
ds^2=a^2(\eta)d\eta^2-a^2(\eta)dx^2-
b^2(\eta)dy^2-b^2(\eta)dz^2.
\label{metric}
\end{eqnarray} 
Here it is assumed that the internal dimensions are frozen
and only the dilaton remains as a dynamical field
whereas the form fields are supposed to have zero
field strength in the background.

The background evolution is then given by \cite{pbb1}
\begin{eqnarray}
a(\eta)=\left[-\frac{\eta}{\eta_1}\right]^{\frac{\alpha}{1-\alpha}},
&\;\;\;\;\;\; &
b(\eta)=\left[-\frac{\eta}{\eta_1}\right]^
{\frac{\beta}{1-\alpha}}\;\; ,
\label{scale}
\end{eqnarray}
and the evolution of the dilaton,
\begin{eqnarray}
\phi(\eta)=\left(\frac{\alpha+2\beta-1}{1-\alpha}
\right)\log\left[-\frac{\eta}{\eta_1}\right]~,
\label{phi}
\end{eqnarray}
with $\alpha$ and $\beta$ satisfying the Kasner 
condition
$$
\alpha^2+2\beta^2=1.
$$
Pre-big-bang inflationary solutions are described by 
$\alpha<0$ and $\beta<0$. Only these will be
considered here.
 
The evolution of the Kalb-Ramond axion field $\si$ defined in Eq.~(2.2)
in Fourier space is determined by \cite{gio}~\cite{ds}
\begin{eqnarray}
\psi_k''+\left[k_L^2+k_T^2\frac{a^2}{b^2}
-\frac{ P''}{ P}
\right]\psi_k=0~,
\label{psi}
\end{eqnarray}
where the canonical field $\psi$ is defined as
$\psi=e^{\phi/2}b\sigma$,
the pump field is  $P=e^{\phi/2}b$,
$k_L$ denotes the modulus of the comoving longitudinal
momentum and $k_T=\sqrt{k_y^2+k_z^2}$ is
the modulus of the transverse momentum.
The anisotropy of the background space-time is 
reflected in the asymmetry between the longitudinal and 
transverse momenta.
For comparison, the case of axion production in an
isotropic space-time has been discussed in
Sec.~IIIB.

Here also we assume that there is an instantaneous transition from the
end of the dilaton phase at $\eta=-\eta_1$ to the
radiation-dominated FLRW post big-bang era.

The aim  is to calculate the spectral energy density
of the axionic inhomogeneities $(d\rho_{\sigma}/
d\log\omega)$ as they re-enter the horizon
during the isotropic radiation-dominated era, after
being amplified during the anisotropic dilaton-dominated epoch.
Inserting the expressions for the scale factors (cf.~(\ref{scale}))
and the dilaton (cf.~(\ref{phi})) in Eq.~(\ref{psi})
one obtains~\cite{gio} \cite{ds}
\begin{eqnarray}
\psi_k''+\left[k_L^2+k_T^2\left(-\frac{\eta}{\eta_1}\right)^{\gamma}
-\frac{\mu^2-1/4}{\eta^2}
\right]\psi_k=0~,
\label{eq.13}
\end{eqnarray}
where
\begin{eqnarray}
\gamma&=&\frac{2(\alpha-\beta)}{1-\alpha}
\;\;\;\;\; 
p=\frac{\alpha+4\beta-1}{2(1- \alpha)}~,\nonumber\\
2\mu&=&\mid 2p-1\mid \;
=2-\frac{4\beta}{1-\alpha}.\nonumber
\end{eqnarray}
We first consider the case $\ga<0$. Then, the term $k_L^2$ always
dominates the bracket in Eq.~(\ref{eq.13}) at very early time $\eta\ra
-\infty$. There is no known exact analytic solution of
Eq.~(\ref{eq.13}). Therefore to make progress
two limiting cases will be discussed \cite{ds}:

\noindent
{\bf Case (I):} The modulus of the longitudinal
momentum, $k_L$, always dominates until 
$\eta^2<1/k_L^2$ at which point the $1/\eta^2$-term comes to dominate. 
This is equivalent to the condition
\be
k_T<k_L(k_1/k_L)^{-\gamma/2}~.
\ee
\noindent
{\bf Case (II):} At some conformal time $\eta=\eta_T<-\eta_1$
the modulus of the transverse momentum, $k_T$,
comes to dominate over $k_L$, but the mode is still
well within the horizon, i.e.
$\sqrt{k_L^2+k_T^2(-\eta_T/\eta_1)^\gamma}
>\eta_T^{-2}$.  Then Eq.~(\ref{eq.13}) implies
\be
\eta_T=-\eta_1\left(\frac{k_L}{k_T}
\right)^{2/\gamma}~.
\ee
Let us examine each of these two cases separately.

\noindent
{\bf Case (I):}
Neglecting the $k_T^2$-term leads to a Bessel equation
which during the pre-big-bang era has the solution,
for $\eta\leq -\eta_1$
\be
\psi_k^{PBB}(k,\eta)=\sqrt{\frac{|k_L\eta|}
{k_L}}H_{\mu}^{(2)}(k_L\eta)~.
\label{sol1}
\ee
During the radiation-dominated FLRW post-big-bang stage,
the solution, for $\eta\geq-\eta_1$, is
\begin{eqnarray}
\psi_k^{RD}=\frac{1}{\sqrt{k}}
\left[c_{+}e^{-ik(\eta+\eta_1)}
+c_{-}e^{ik(\eta+\eta_1)}
\right].
\label{sol2}
\end{eqnarray}
Matching the two solutions Eq.~(\ref{sol1})
and Eq.~(\ref{sol2}) determines the frequency mixing
coefficient $c_{-}$ which allows to calculate the occupation
numbers of produced axions.
The spectral energy density of the produced axions is 
\begin{eqnarray}
\rho_L=\frac{d\rho_{\sigma}}{d\log\omega}
\simeq\frac{\omega^4}{\pi^2}|c_{-}|^2~,
\end{eqnarray} 
which with 
\begin{eqnarray}
c_{-}=-\frac{1}{2\pi}\sqrt{\frac{1}{(k\eta_1)
(k_{L}\eta_1)^{2\mu}}}
\end{eqnarray}
yields to
\begin{eqnarray}
\rho_L(\omega,s)\simeq\frac{\omega_1^4}{2\pi^3}
s^{-2\mu}\left(\frac{\omega}{\omega_1}\right)^{3-2\mu}~,
\end{eqnarray}
where $s=k_L/k$.
Note that in the special case $\mu=3/2$,
corresponding to $\alpha=-7/9$, $\beta=-4/9$,
a flat spectrum is obtained.

\noindent
{\bf Case (II):}
Assuming that the $k_T$-term comes to dominate before
the perturbation becomes super-horizon Eq.~(\ref{eq.13})
may be approximated by
\begin{eqnarray}
\psi_k''+\left(k_L^2
+k_T^2\left[-\frac{\eta}{\eta_1}\right]^\gamma
\right)\psi_k=0.
\label{eq.25}
\end{eqnarray}
An approximate solution of Eq.~(\ref{eq.25}) is
\begin{eqnarray}
\psi\simeq
\frac{\exp(i\eta\sqrt{k_L^2+q^2(-\eta/\eta_1)^\gamma
k_T^2})}
{\sqrt{\pi/2}[k_L^2+(-\eta/\eta_1)^\gamma k_T^2]^{1/4}}
\label{eq.26}
\end{eqnarray}
where $q=1/(1+\gamma/2)=(1-\alpha)/(1-\beta)$.
This represents the in-coming vacuum solution.

In Fig.~3, we present numerical solutions of 
Eq.~(\ref{eq.25}) and compare them with
the approximate analytical solution Eq.~(\ref{eq.26}) for different values
of $s$ and $\ga$. 
\begin{figure}
\centerline{\epsfxsize=2.5in\epsfbox{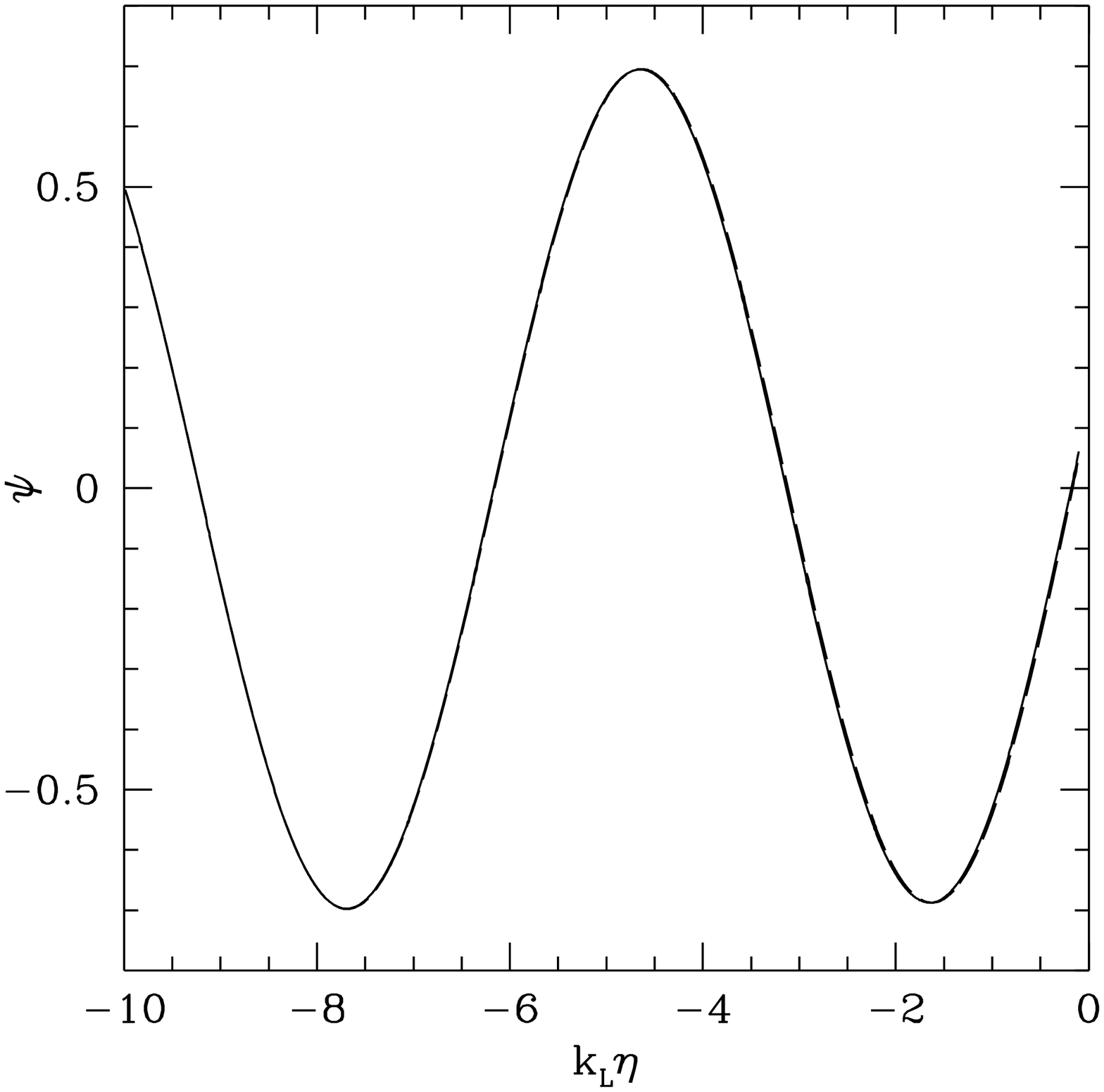} \hspace{1cm}
              \epsfxsize=2.5in\epsfbox{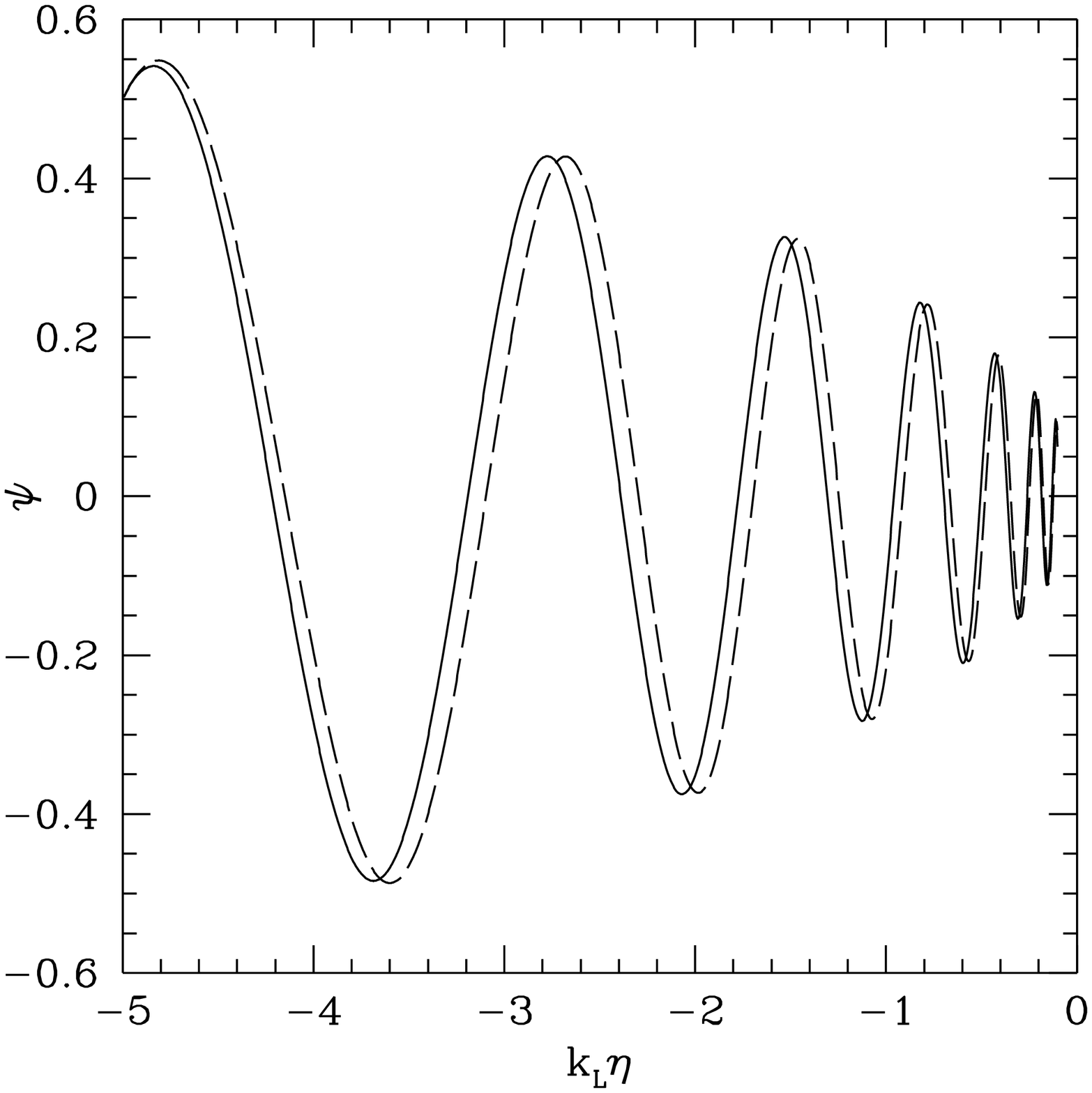}}
\caption{Comparison of the numerical and approximate
solution for $s=0.93$, $\gamma=-0.5$ (left) and  $s=0.1$,
$\gamma=-1.9$ (right). The solid line
represents the numerical solution and the dashed one the
approximate solution.}
\label{fig3}
\end{figure}
As the comparison shows the approximate
analytic solution Eq.~(\ref{eq.26}) is very good
for small as well as large values of $s$
(cf. Fig.~\ref{fig3}).

\begin{figure}[ht]
\centerline{\epsfxsize=2.2in  \epsfbox{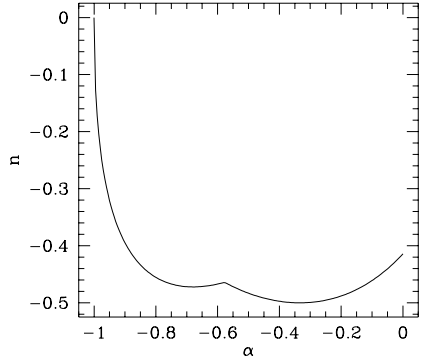}}
\caption{The spectral index $n$ of the axion perturbations  is shown
as a function of the Kasner exponent $\alpha$. Except in the
one-dimensional limit $\al\ra -1$, the spectral index is always very
close to the isotropic result, the value for $\al=1/\sqrt{3}$.}
\label{nax}
\end{figure}

At conformal time $\eta=\eta_T$, the 
transverse momentum $k_T$ becomes dominant
over the $k_L$ term and finally
the $\eta^{-2}$ dominates. After time $\eta=\eta_T$ Eq.~(\ref{eq.13}) 
can be approximated by
\begin{eqnarray}
\psi_k''+\left[k_T^2\left(-\frac{\eta}{\eta_1}\right)^{\gamma}
-\frac{\mu^2-1/4}{\eta^2}~,
\right]\psi_k=0
\end{eqnarray}
which has the general solution
\begin{eqnarray}
\psi_k&=&c_T^{(1)}\sqrt{|k_T\eta|}
H_{\mu q}^{(1)}
\left(|k_T\eta| q
\left[\frac{-\eta}{\eta_1}\right]^{\gamma/2}
\right)
\nonumber\\
&&
-ic_T^{(2)}\sqrt{|k_T\eta|}
H_{\mu q}^{(2)}
\left(|k_T\eta| q
\left[\frac{-\eta}{\eta_1}\right]^{\gamma/2}~
\right)~,
\label{psi_pbb}
\end{eqnarray}
$H_{\mu q}^{(1)}$ and  $H_{\mu q}^{(2)}$
are Hankel functions of the first and second kind of order
$\mu q$. For large $k_T| \eta|$ the second term 
just corresponds to the approximate solution, Eq.~(\ref{eq.26}).
Therefore matching these solutions, the coefficients are
found to be
\be
c_T^{(1)}=0\;\;\; , \;\;\;
c_T^{(2)}=\frac{i}{\sqrt{k_T}}~.
\label{c_T}
\ee
For super-horizon perturbations ($|k_T\eta_1|\ll 1$), 
matching the physical fields at the transition time
$\eta=-\eta_1$, for $|k_T\eta_1|\ll 1$ (i.e., from the end of dilaton-driven
era to the beginning of the radiation-dominated post-big-bang universe) one obtains
the Bogoliubov coefficient $c_{-}$:
\begin{eqnarray}
|c_{-}|^2=
\left[\frac{\Gamma^2(\mu q)}{4\pi^2}
2^{2\mu q}\left(\frac{3}{2}-\mu q\right)^2\right]
\left(\frac{k_T}{k_L}\right)^{-2\mu q}
s^{-2\mu q}
\left(\frac{\omega}{\omega_1}\right)^{-1-2\mu q}~.
\end{eqnarray}
The energy density of the produced Kalb-Ramond axions,
in the case where the transverse momentum $k_T$ is dominant,
is
\be
\rho_T(\omega,s)=
\left[\frac{\Gamma^2(\mu q)}{4\pi^2}
2^{2\mu q}\left(\frac{3}{2}-\mu q\right)^2\right]
\frac{1}{\pi^2}\left(\frac{k}{k_T}\right)
^{2\mu q}\omega_1^{1+2\mu q}
\omega^{3-2\mu q}~.
\ee
Therefore, in summary, we obtain
\begin{eqnarray}
\rho(\omega,s)\simeq {\om_1^4\over 2\pi^3}
\left\{\begin{array}{l}
  s^{-2\mu} \left({\om\over \om_1}\right)^{3-2\mu }
\;\;\;\;\;	\mbox{ if }  k_T<k_L(k_1/k_L)^{-\g/2} \\
(1-s^2)^{-\mu q} \left({\om\over \om_1}\right)^{3-2\mu q} 
\;\;\;\;\;  ~\mbox{ else.}
\end{array}\right.
\end{eqnarray}
This can be expressed in terms of
$\Omega_{\gamma}(\eta)=(H_1/H)^2(a_1/a)^4$, i.e. of the fraction of
critical energy density in radiation at a given time $\eta$, and of
$g_1=H_1/M_{\rm Pl}$, the transition scale in units of the Planck mass
$M_{\rm Pl}$, as
\begin{eqnarray}
\Omega_\sigma(\omega,s,\eta) &\simeq& g_1^2\Omega_{\gamma}(\eta)
\nonumber \\
 && \times\left\{\begin{array}{l}
(1-s^2)^{-\mu q} \left({\om\over \om_1}\right)^{3-2\mu q}
\;\;\;\;\;\mbox{ if }  s\leq s_c(\omega) \\
 s^{-2\mu} \left({\om\over \om_1}\right)^{3-2\mu }
\;\;\;\;\;	\mbox{ if } s\geq s_c(\omega) \\
\end{array}\right.
\end{eqnarray}
For a given value of $\omega$, the parameter $s_c$ is determined
by the equation
\begin{eqnarray}
\sqrt{1-s_c^2}=s_c^{1+\gamma/2}
\left(\frac{\omega}{\omega_1}\right)^{\gamma/2}.
\label{s_c}
\end{eqnarray}
To estimate the total energy density per logarithmic frequency
value one has to integrate the axion density $\Omega_\sigma
(\omega,\eta,s)$ over $s$. Using $d^3k=4\pi k^2ds\wedge dk$
results in the following expression for $\Omega_\sigma(\omega,\eta)$:
\begin{eqnarray}
\Omega_\sigma(\omega,\eta)&=&\int\Omega_\sigma(\omega,\eta,s)ds
\nonumber\\
&\simeq& g_1^2\Omega_{\gamma}
\left[
 \left(\frac{\omega}{\omega_1}\right)^{3-2\mu q}
 \int_0^{s_{c}(\omega)}(1-s^{2})^{-\mu q}ds
+\left(\frac{\omega}{\omega_{1}}\right)^{3-2\mu}
 \int_{s_{c}(\omega)}^{1}s^{-2\mu}ds
\right].\label{OM}
\end{eqnarray}
Using that $\omega<\omega_1$ the value of $s_c$
can be approximated in the two cases $\gamma<0$ and $
\gamma>0$ as follows,
\begin{eqnarray}
s_c&\simeq&\left(\frac{\omega}{\omega_1}\right)^{q-1}
\;\;\;\;\;\mbox{ if } \gamma<0 \label{sc_app1} \\
1-s_c^2&\simeq&\left(\frac{\omega}{\omega_1}\right)^{\frac{2}{q}-2}
\;\;\;\;\;\mbox{ if } \gamma>0 \label{sc_app2}
\end{eqnarray}
with $q=1/(1+\gamma/2)$. Strictly speaking, our analysis is only
valid for $\ga<0$, but the result remains correct also for $\ga>0$,
as one can  easily check using our approximation Eq.~(\ref{eq.26}), which
in this case then describes the in-coming vacuum solution.
Thus, the integrals in Eq.~(\ref{OM}) can be approximated, 
to give  
\begin{eqnarray}
\Omega_\sigma(\omega,\eta)\simeq
g_{1}^2\Omega_{\gamma}(\eta)\left(\frac{\omega}{\omega_1}
\right)^{n}~,
\end{eqnarray}
where
\begin{eqnarray}
n&=&2+q-2\mu q=\frac{1+\alpha+2\beta}{1-\beta}~~~
\mbox{ if } \alpha<\beta \nonumber\\
n&=&1+\frac{2}{q}-2\mu=\frac{1+\alpha+2\beta}{1-\alpha}~~~
\mbox{ if } \alpha>\beta~. \nonumber
\end{eqnarray}
Since, $\alpha^2+2\beta^2=1$ and for PBB inflation
$\alpha, \beta\leq 0$, it follows that $	\alpha+2\beta\leq -1$.
This implies that the spectrum is generically red and only the
degenerate case with two static and one inflating dimensions,
($\alpha=-1,\beta=0$) yields a flat spectrum. The spectral
index is relatively close to the isotropic value, $n_{iso}
=3-2\sqrt{3}\sim -0.46$, for all reasonable values of $\al$ and $\beta$
(see Fig.~\ref{nax}).
In conclusion, the anisotropic expansion has little influence on the
overall axion production, while as we have seen in the previous section,
extra dimensions can yield generically a flat spectrum of axions.

\subsection{Photon production}

An interesting aspect of photon production in anisotropic space-times
is that even without the coupling of the dilaton to the Maxwell part
of the action, there is photon production since anisotropic
space-times are not conformally flat \cite{b1mag}. Again, we restrict our
analysis to the axi-symmetric case.  In the string frame Maxwell's
equations read
\begin{eqnarray}
\partial_{\mu}(e^{-\phi}\sqrt{-g}F^{\mu\nu})=0~, \label{F1}\\
\partial_{\mu}(\sqrt{-g}\star F^{\mu\nu})=0~, \label{F2}
\end{eqnarray}
Setting $F_{\mu\nu}=\partial_{\mu}A_{\nu}-\partial_{\nu}A_{\mu}$, 
where $A_\mu$ is the electromagnetic gauge potential, we solve
Eq.~(\ref{F2}) identically. For a metric $ds^2=-dt^2+a_i^2(dx^i)^2$
Eq.~(\ref{F1}) leads in momentum-space to:
\begin{eqnarray}
\sum_{i=1}^{3}a_i^{-2}ik_i[\partial_0 A_i
-ik_iA_0]&=&0~,
\label{constr}\\
-\partial_0[e^{-\phi}\sqrt{-g}a_m^{-2}
(\partial_0A_m-ik_mA_0)]
+e^{-\phi}\sqrt{-g}a_m^{-2}
\sum_{n=1}^{3}a_n^{-2}ik_n
[ik_nA_m-ik_mA_n]&=&0~,
\label{box}
\end{eqnarray}
where $m=1,2,3$ (no sum).

We impose the gauge conditions $A_{0}\equiv 0$
and  $\underline{k}.\underline{A}=0
\Rightarrow\sum_i g^{ii}k_iA_i=0
\Rightarrow\sum_i a_i^{-2}k_i A_i=0$ which corresponds to the radiation gauge.
 Using the axi-symmetric metric Eq.~(\ref{metric}), this gauge condition
together with Eq.~(\ref{constr}) imply
the constraint,
\begin{eqnarray}
A_L\equiv 0 &\;\;\;\;{\rm or}\;\;\;\;& k_L\equiv 0~,
\end{eqnarray}
where the index $L$ denotes the longitudinal direction. The gauge
condition then reduces to $\vec{k_T}\cd\vec{A_T}=0$, where
$\vec{v_T}$ denotes the two-dimensional vector in the $(y-z)$
plane. $\vec{A_T}$ hence has one degree of freedom, normal to
$\vec{k_T}$, which we simply denote by $A_T$.
Due to the breaking of spherical symmetry the longitudinal ($A_L$) and
transverse ($A_T$) degrees of freedom obey different equations of motion.
Introducing the canonical fields
$\psi_L\equiv e^{-\phi/2}(b/a)A_L$
and $\psi_T\equiv e^{-\phi/2}A_T$, 
Eq.~(\ref{box}) leads to 
\begin{eqnarray}
\psi_{L}''+\left[k_L^2+\left(-\frac{\eta}{\eta_1}
\right)^{\gamma}k_{T}^2-\frac{\lambda_L}{\eta^2}\right]
\psi_L&=&0~,\\
\psi_{T}''+\left[k_L^2+\left(-\frac{\eta}{\eta_1}
\right)^{\gamma}k_{T}^2-\frac{\lambda_T}{\eta^2}\right]
\psi_T&=&0~,
\end{eqnarray}
where  
\bea
 \gamma &\equiv& {2(\alpha-\beta)\over (1-\alpha)}~~,~~
\lambda_L\equiv {(3\alpha-1)(1+\alpha)\over 4(1-\alpha)^2} \nonumber \\
\lambda_T &\equiv& {(-\alpha+2\beta+1)(\alpha+2\beta-1)
\over 4(1-\alpha)^2}~.
\eea
There are two cases to discuss. The above set of equations
has to be solved for either (a) $\psi_L\equiv 0$ or 
(b) $k_L\equiv 0$.
Case (a) reduces the system to just one independent
equation. Therefore, in this case, we can simply adapt the discussion
of the previous section on axion production in an anisotropic background.
In case (b) there are two independent equations. However, since
$k_L\equiv 0$ they reduce to Bessel equations which are
exactly solvable.

\begin{figure}[ht]
\centerline{\epsfxsize=2.2in  \epsfbox{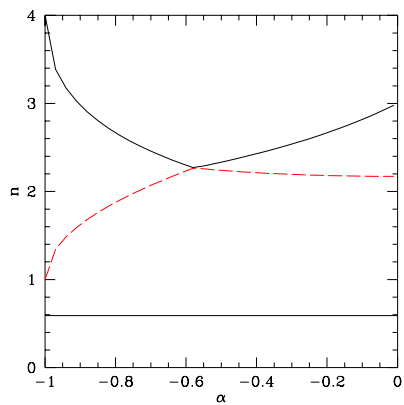}}
\caption{The spectral index $n$ for photon production in an
anisotropic pre-big-bang (case~(a) solid, case~(b) dashed) and is
shown as a function of  the Kasner exponent $\alpha$. To obtain a strong 
enough magnetic field for successive amplification
by a galactic dynamo mechanism $n$ has to lie below
the line $n\sim 0.59$.}
\label{fig5}
\end{figure}

\begin{itemize}
\item {\bf Case (a):} $\Psi_L\equiv 0$

\begin{itemize}
\item{\bf Case (I):} 
If the longitudinal momentum ($k_L$) dominates as long as the
perturbation is sub-horizon, the spectral energy density in the
radiation era is obtained exactly like in the axionic case:
\begin{eqnarray}
\rho_L(\omega,s)\simeq 2\frac{\omega_1^4}{2\pi^3}
s^{-2\mu_T}
\left(\frac{\omega}{\omega_1}\right)^{3-2\mu_T}~,
\end{eqnarray}
where $\mu_T\equiv(\frac{1}{4}+\lambda_T)^{\frac{1}{2}}=
\left|\beta/(1-\alpha)\right|.$
A flat spectrum is recovered for $\mu_T=3/2$.
 Using the Kasner constraint this corresponds to a positive value of
$\al$, namely
$\alpha=7/11$, $\beta=\pm 6/11$. Hence 
in this case, PBB inflation only takes place in the transverse but not
in the longitudinal direction. 

\item {\bf Case (II)}
If the transverse momentum ($k_T$) comes to dominate on sub-horizon scales,
the in-coming vacuum solution is approximately given
by Eq.~(\ref{eq.26}). The solution for super-horizon modes
during the dilaton-driven inflationary stage
is again given by Eqs.~(\ref{psi_pbb}),(\ref{c_T}) with $\mu=\mu_T$. 
The matching of the gauge potential
and its first derivatives at the transition from 
the dilaton-driven era to the radiation-dominated FLRW universe at
$\eta=-\eta_1$ determines the Bogoliubov coefficient $c_{-}$.
Recalling that $A_T=e^{\phi/2}\psi_T$, 
it is found that for $|k_T\eta_{1}|\ll 1$,
\be
|c_{-}|^2=\frac{1}{4\pi^2}\Gamma(\mu_T q)^2
\left(\frac{q}{2}\right)^{-2\mu_T q}
(k_T\eta_1)^{-2\mu_T q}(k\eta_1)^{-1}
\left[\left(\mu_T-\frac{\beta}{1-\alpha}\right)^2
+{\cal O}(k\eta_1)^2\right]~,
\label{c-}
\ee
where $q=(1-\alpha)/(1-\beta)$ and 
$\mu_T=\left|\beta/(1-\alpha)\right|$.
The first term in the last square bracket vanishes
for  positive $\beta$. However, only
solutions with $\beta<0$ are really of interest here
since they describe PBB inflationary expansion.
In summary, for $\beta<0$ the spectral energy density of the 
produced photons is 
\begin{eqnarray}
\rho_T(\omega,s)\simeq
2\frac{\mu_T^2}{\pi^4}\omega_1^4\left[\Gamma(\mu_T q)\right]^2
\left(\frac{q}{2}\right)^{-2\mu_T q}
(1-s^2)^{-\mu_T q}
\left(\frac{\omega}{\omega_1}\right)^{3-2\mu_T q}~.
\end{eqnarray}
%
Thus the density parameter $\Omega_{em}$
is 
\begin{eqnarray}
\Omega_{em}(\omega,s,\eta)\simeq 2 g_1^2\Omega_{\gamma}(\eta)
\left\{\begin{array}{ll}
(1-s^2)^{-\mu_T q} \left({\om\over \om_1}\right)^{3-2\mu_Tq} &
\mbox{if }  s\leq s_c(\omega) \\
 s^{-2\mu_T} \left({\om\over \om_1}\right)^{3-2\mu_T} &
\mbox{if } s\geq s_c(\omega)~. \\
\end{array}\right.
\end{eqnarray}

In order to estimate the total energy density per 
logarithmic frequency interval, $\Omega_{em}(\omega,s,\eta)$
has to be integrated over $s$. The value of $s_c$ is determined
as before by Eq.~(\ref{s_c}). Carrying out the integration,
making use of Eqs.~(\ref{sc_app1}) and (\ref{sc_app2}),
the density parameter of the produced electromagnetic
spectrum reads
\begin{eqnarray}
\Omega_{em}(\omega,\eta)\sim g_1^2\Omega_{\gamma}(\eta)
\left(\frac{\omega}{\omega_1}\right)^{n}~,
\end{eqnarray}
with the spectral index $n$, 
\begin{eqnarray}
n=\left\{\begin{array}{l}
    2+q-2\mu_T q=
    \frac{3-\alpha}{1-\beta} \;\;\;\;\;\mbox{ if } \alpha<\beta\\
    1+\frac{2}{q}-2\mu_T=
    \frac{3-\alpha}{1-\alpha} \;\;\;\;\;\mbox{ if } \alpha>\beta~.\\
\end{array}\right.
\end{eqnarray}
In contrast to the axion case, the photon spectrum is always blue.
For $\alpha=\beta=-1/\sqrt{3}$ the isotropic
spectral index, $n_{iso}=4-\sqrt{3}$ is recovered~\cite{MG}.
In Fig.~\ref{fig5} the spectral index $n$ is shown as a function of $\al$.
\end{itemize}

\item {\bf Case (b)}: $k_L\equiv 0$

This case is very particular  since it describes the
production of photons with wavenumber confined to the symmetry plane.

Since $\psi_L\not\equiv 0$, there are two independent equations
\begin{eqnarray}
\psi_{L}''+\left[\left(-\frac{\eta}{\eta_1}
\right)^{\gamma}k_{T}^2-\frac{\lambda_L}{\eta^2}\right]
\psi_L&=&0~,\\
\psi_{T}''+\left[\left(-\frac{\eta}{\eta_1}
\right)^{\gamma}k_{T}^2-\frac{\lambda_T}{\eta^2}\right]
\psi_T&=&0~.
\end{eqnarray}
These equations can be solved in terms of Bessel
functions. However, the overall discussion 
follows closely that of case (a) (II). The in-coming vacuum
solution is again given by Eq.~(\ref{eq.26}) which in this
case is an exact solution during the pre-big-bang phase. The matching
procedure is the same for $A_{T}$, but note that the solution for 
$\psi$ in the radiation-dominated era is also just a function of
$k_T=k$.
Performing a similar calculation as before for $A_L$, the Bogoliubov
coefficient for $A_L$--photons is 
\begin{eqnarray}
|c^{(L)}_{-}|^2 &=&\frac{1}{4\pi^2}\left[\Gamma(\mu_L q)\right]^2 \nonumber\\
&& \times\left(\frac{q}{2}\right)^{-2\mu_L q}
\left(k_T\eta_1\right)^{-2\mu_L q-1}
\left[\left(-\frac{\alpha}{1-\alpha}+\mu_L\right)^2
+{\cal O}(k_T\eta_1)^2\right]~,
\end{eqnarray}
where $\lambda_L\equiv\mu^2_L-1/4$ implies
$\mu_L=\left|\alpha/(1-\alpha)\right|$.

The spectral energy density, keeping in mind that this result holds
just for $s=k_L/k=0$, is 
\be
\rho \simeq\frac{\omega^4}{\pi^2}\left(|c_{-}^{(L)}|^2 +
	|c_{-}^{(T)}|^2\right)\delta(s) ~,
\ee
where $|c_{-}^{(T)}|^2$ is given by Eq.~(\ref{c-}) with
$k$ replaced by $k_T$. Integration over directions then  yields the
density parameter  
\begin{eqnarray}
\Omega_{em}(\omega,\eta)\simeq
g_1^2\Omega_{\gamma}\left(
{\cal N}_L \left(\frac{\omega}{\omega_1}\right)^{m_L} + {\cal N}_T
\left(\frac{\omega}{\omega_1}\right)^{m_T}\right) ~,
\end{eqnarray}
where ${\cal N}_{\bullet}$ collects all the numerical factors of order
unity in  $\rho$  and $|c_{-}^{(\bullet)}|^2$, respectively,
and $m_{\bullet}$ are given by
\begin{eqnarray}
m_L&=&\frac{3+2\alpha-3\beta}{1-\beta} \nonumber\\
m_T&=&\frac{3-\beta}{1-\beta}.
\end{eqnarray}
The ``effective spectral index'', the smaller of the two,
$m_*=\min(m_L,m_T)$, is indicated in Fig.~\ref{fig5}. Again, we always
obtain blue spectra, if all dimensions are expanding, $\al,\beta <0$.
\end{itemize}

In order to discuss the production of primordial
magnetic fields it is useful to introduce a
parameter $r(\omega)$ defined by \cite{10}
\begin{eqnarray}
r(\omega)\equiv\frac{\Omega_{em}}{\Omega_{\gamma}}.
\end{eqnarray}
Galaxies are endowed with magnetic fields of typical strength of order
$10^{-6}$ G, coherent on a comoving scale of $\lambda_G\sim 10$ kpc.
Assuming the existence of some kind of galactic dynamo $r$ needs to be
of order $r(\omega_G)\geq 10^{-34}$~\cite{10}.

This implies a constraint on the spectral index.  With
$\omega_G\simeq(10^{-2}{\rm Mpc})^{-1} \simeq 10^{-36}{\rm GeV}$ and
$\omega_1\simeq H_1 =g_1 M_{Pl}\simeq10^{16}-10^{18}$GeV, where
$g_1\simeq 10^{-3}-10^{-1}$, this requires $n<0.59$.
This cannot be achieved in this anisotropic PBB model (cf.
Fig.~\ref{fig5}).  Furthermore, the spectral index is minimal for the
photons produced on an isotropic background, or for $k_L=0$.

Clearly, in the isotropic case the contributions from cases (a) and
(b) coincide. Note however that in the anisotropic case photon
production in the plane $k_L=0$ dominates ($m_*<n$   , flatter spectra)
over the polarized photons with $A_L=0$. The $k_L=0$ photons are
specific to the anisotropic, not conformally flat case and do lead to
an enhancement of photon production in this case. Nevertheless, 
the effect of global anisotropies is not sufficient to produce the
necessary primordial magnetic fields. Once again allowing for evolving
extra dimensions or a long intermediate string phase, one  can push the
spectral index sufficiently down to create strong enough magnetic seed fields
\cite{7}.

\section{Conclusions}

We have analyzed some phenomenological aspects induced by particle
production in the PBB scenario, which is a particular cosmological
model inspired by the duality properties of string theory. Assuming
that the transition from the pre- to the post-big-bang era will not
affect the observational consequences of the PBB model, we compare the
theoretical observables with current observational data. In doing so,
we provide a test for string theory as a fundamental theory and fix
some of the parameters of the PBB model.

We first considered a $D$-dimensional space-time, containing a
four-dimensional homogeneous and isotropic external metric and a
$(D-4)$-dimensional compactified internal metric, with vanishing axion
contribution. We study the evolution of perturbations that may be
generated by the parametric amplification of vacuum fluctuations as
the universe goes from the pre- to the post-big-bang era. We study
scalar and tensor metric perturbations. We find that scalar
perturbations have generically red spectra while tensor perturbations
have blue spectra. A 
robust prediction of the PBB model is $n_{\rm T}=3$, while standard
inflationary models require $n_{\rm T}<0$. Dilaton and moduli field
perturbations  in the standard pre-big-bang model may
  acquire red spectra with a spectral index $n=0$.
This is of course very different from a scale-invariant
Harrison-Zel'dovich spectrum with $n=1$ and leads to large
perturbations on large, cosmologically relevant scales during the
radiation era. Therefore the ordinary pre-big-bang model is in
conflict with observations if the $n=0$ spectrum is
inherited. However, if one includes an exponential 
dilaton potential, the dilaton and moduli perturbations can induce scale
invariant adiabatic perturbations in the radiation era like ordinary
inflation which are in perfect accordance with present observations
e.g. of CMB anisotropies. The blue spectra of the tensor modes
(gravity waves), which are
normalized to $g_1^2$ at the high frequency end, decay rapidly towards
longer wavelengths and are completely negligible on cosmological scales.

Even if a field does not contribute to the background evolution,
quantum fluctuations cannot be neglected. The induced energy density
perturbation is then of second order in the field perturbation, 
but it can lead to appreciable perturbations in space-time
geometry. Perturbations of fields which do not contribute to the
background are referred to as ``seeds''. We analyze electromagnetic and
Kalb-Ramond axion seeds, and study their r\^ole for the origin of
primordial galactic magnetic fields, the large-scale structure and CMB
anisotropies. We compute the stochastic fluctuations of the
energy-momentum tensor of the seeds and determine their contribution
to the multipole expansion of the temperature anisotropy.  We find
that electromagnetic perturbations lead to a ``blue'' power spectrum
whose amplitude is fixed at the string scale; on larger scales, it
decays too fast to produce primordial magnetic fields which may be
amplified to the presently observed values.  However, since the
contribution of electromagnetic perturbations to the large-scale anisotropy 
is negligible, the COBE normalization does not impose any constraints to the 
production of seeds for galactic magnetic fields. Kalb-Ramond axions, which are
either massless or  have a mass up to $100 MeV$ can lead to a flat
or slightly blue spectrum, in reasonable agreement with current
data. The actual value of the axion spectral index depends on the rate
of contraction of the internal dimensions during the pre-big-bang era.

We also consider four-dimensional spatially flat anisotropic PBB
cosmological models. Computing the energy spectra for massless
Kalb-Ramond axions, we find that, when integrated over directions,
this model leads to infra-red divergent spectra, as in the case of a
four-dimensional spatially flat isotropic model. Similarly, analyzing
photon production in this background, we find that the obtained blue
spectra do not differ significantly from the isotropic
case. Therefore, also the four-dimensional anisotropic pre-big-bang
models suffer from the unphysical red axion spectrum and the blue
photon spectrum found in the four-dimensional isotropic pre-big-bang models.

The analysis presented here is valid for modes which exit the horizon
before the string phase. However, if the intermediate string phase is
sufficiently long, the spectra can be affected on sufficiently large scale
which may be of interest for cosmology. Some examples of spectra from
significantly long string phase have been explored in
Refs.~\cite{axionmass,7,Buon}).
\vspace{0.2cm}\\
{\bf Acknowledgment}\\
We have benefited from discussions with Alessandra Buonanno, Ed
Copeland, Gabriele Veneziano and Filippo Vernizzi. The work of
R.D. and K.E.K. is supported by the Swiss NSF.

\end{document}